# Dynamic Conditional Correlation between Electricity and Stock markets during the Financial Crisis in Greece.

**Panagiotis G. Papaioannou[1], George P. Papaioannou[2,3], Kostas Siettos[1], Akylas Stratigakos[4] , Christos Dikaiakos[2],**

[1]Applied Mathematics and Physical Sciences, National Technical University of Athens, Greece; pgp2ntua@central.ntua.gr    ksiet@mail.ntua.gr

[2]Research, Technology & Development Department, Independent Power Transmission Operator (IPTO) S.A., 89 Dyrrachiou & Kifisou Str. Gr, 104 43 Athens, Greece; g.papaioannou@admie.gr; c.dikaiakos@admie.gr

[3]Center for Research and Applications in Nonlinear Systems (CRANS), Department of Mathematics, University of Patras, Patras 26 500, Greece

[4]Department of Electrical and Computer Engineering, University of Patras, Patras 26500, Greece; akylas.strat@hotmail.gr

## ABSTRACT

Liberalization of electricity markets has increasingly created the need for understanding the volatility and correlation structure between **electricity** and **financial** markets. This work reveals the existence of structural changes in correlation patterns among these two markets and links the changes to both fundamentals and regulatory conditions prevailing in the markets, as well as the current European financial crisis. We apply a Dynamic Conditional Correlation (DCC) GARCH model to a set of market's fundamental variables and Greece's financial market and microeconomic indexes to study their interaction. Emphasis is given on the period of severe financial crisis of the Country to understand "contagion" and volatility spillover between these two markets.

## 1.  Introduction and literature review

In the financial and Commodity markets, conditional volatility models have found an extensive application. However the studies focusing on modeling the spillover of price conditional volatility between financial, energy (commodity) and wholesale electricity markets in Europe are very few.

The transmission of price volatilities between two natural gas markets, the British and Belgium ones, is investigated by Bermejo-Apricio et al, (2008). They applied CARCH (1,1) and EGARCH (1,1) for the univariate case and a DCC and BEKK for the bivariate case, on deseasonalized daily prices of NBP and Zeebrugge Hubs. They took also into consideration the Interconnector gas pipeline's used capacity as an exogenous variable for the conditional variance. Their study has shown the existence of an **inverse leverage** effect for the Zeebrugge and NBP prices i.e. large price increases (positive shock) increase the conditional volatility more than large price drops (negative shock). The main conclusion in their paper is that the Interconnector gas pipeline impacts strongly the conditional variance of NBP and Zeebrugge, resulting in an increase of the volatility linkage between the two markets when 50% or more of the pipe-line's total capacity is used.

The interaction between gas spot prices at Zeebrugge, one month-ahead Brent Oil Prices and temperature, for period 2000-2005, is examined in the work of Regnard and Zokoian (2011). They used a Vector Error Correction Model (VECM) to investigate the joint dynamics of the three variables and found (using Johansen's approach) evidence of a cointegrating linkage between the three variables. Also, using an asymmetric Constant Conditional Correlation (A-CCC) model and multivariate GARCH have shown that volatilities of the three series are dependent on their own lagged volatilities. They found significant cross-effects in the conditional correlation matrix. They also examine the influence of 3 different temperature regimes on the conditional variance (low temperature regime positive shocks increase the conditional variance, while the Zeebrugge price's volatility is increased due to negative shocks originating from high temperature regime).

The interaction between Brent Oil and NBP spot price returns is estimated by Asche et al. (2009), conducting a multivariate GARCH and a BEEK model. They show that prior to 2003 (a year corresponding to a breakdown), there is not any impact of shocks occurred in the oil (gas) market on the conditional variance of gas (oil). They argue that a possible explanation of the impacts of oil price shocks on the volatility of gas prices is the small or limited available capacity of the European gas market infrastructure as well as the enhanced levels of maturity and liquidity of the European NG spot market.

The volatility spillovers between the $CO_2$, Brent Oil and gas markets in Europe, is the main theme of the paper by Chevallier (2012) and Mansanet – Bataller and Soriano (2009). In both studies, data of $CO_2$ price series daily futures for the December 2008 contract are used. Daily NYMEX Crude Oil futures and Zeebrugge next month contract prices are used in Chevallier (2012) paper, while front month prices for Brent Oil and NG are used in the Mansanet-Bataller-Soriano's (2009) paper. Trivariate multivariate GARCH models, namely the CCC, DCC and the BEEK model were used to "capture" the volatility spillovers in Chevallier's (2012) work, while a BEEK model is used in the case of the other paper. The DCC model used in Chevallier's work shows that the conditional correlation between Oil and NG is from -0.3 to 0.3 and for NG and $CO_2$ is from -0.2 to over 0.1.

Koch, N. (2014) has studied the dynamic linkages among Carbon (eua), Commodity (Energy) and financial markets using the Smooth Transition Conditional Correlation (STCC) approach, an extension of DCC model. He calls Oil, gas, coal, electricity, stocks and bonds as **accepted fundamentals.** He used time as transition variable to allow for structural breaks related to institutional changes in the EU ETS. His main conclusion is that correlation depends on market uncertain conditions, reflecting the connection between **Carbon** and **Financial markets** due to common macroeconomic shocks happened over the 2008/09 financial crisis.

The linkage between EUA and Financial markets as described in Koch's work serves as a firm basis on which we build our work. The component "EUA-financial markets linkage" or more precisely the volatility spillover between eua-financial markets, is depicted in the following triplet "causal map"

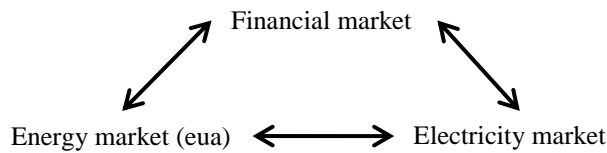

We share Koch's argument on the existence of correlation asymmetries due to time-varying market uncertain conditions and examine in our work here the influences of these conditions on the dynamic conditional correlations during periods of calmness and turmoil in financial markets. This is of particular value in the case of Greece, a State hit heavily by two crisis, the financial one 2008-2009 and the Greek Debt (Sovereignty) crisis started in late 2010.

## 2. Macroeconomic Risk factors and the significance of Volatility spillover or Co-movement

The importance of macroeconomic risk factors in shaping the expectations of the equity, bond and commodity markets, has been "stressed" by Fama and French (1989) and Sadorsky (2002). These factors are assumed in this work to influence Carbon, Energy and Electricity markets (Chevalier, 2009). Thus, we expect the EUA price to fall if there is a prospective economic slow-down, indicated by the macroeconomic indicators. This is a rational expectation since adverse business conditions lower aggregated demand and thus reduce the demand for electricity (load), the generation output, the demand for coal and as a consequence the demand for EUA. The two stock indices ase, Euro stoxx50 and vstoxx (for volatility) are considered, as well as the 10-year Greek Government Bonds as measures for macroeconomic and financial risks in Greece and Europe. The stock indices measure the development of the financial markets and are used to predict the fluctuations of the general economic "climate".

### 2.1 The importance of input fuel prices volatilities and their co-movement with EUA

The operational behavior that links fuel and EUA is the generator's fuel-switching. This is so because a higher gas (coal) price ends up to a higher (lower) eua:

$$ngasUK \uparrow \quad then\ eua \uparrow$$

$$coal \uparrow \quad then\ eua \uparrow$$

This observation is a good theoretical basis for explaining the co-movement or the Dynamic Conditional Correlation between input fuel prices and eua. A producer of electric power uses hydrocarbon fuels and eua as production inputs, so he depends on these "assets". This situation is not the same as in a financial market in which a portfolio manager can diversify his assets portfolio by altering the (percentage) share of the assets, in order to protect the value of the portfolio from price changes (hedging). The power producer is exposed to changes in prices in electricity, energy (commodity) and EUA markets. Therefore, the **risk-averse** Power Plant Owner (producer) has to operate in forward (futures) markets for hedging his profits against the risk of unpredictable and unfavorable price volatility. In other words he tries to lock in a given profit based on a given (assumed) **marginal** generation cost.

However, the key variables in a futures market are the price volatility of an "asset" (input fuel, eua etc.) and its co-movement with other relevant asset's price. This co-movement is measured by its **conditional covariance** or **correlation** price volatility is usually expressed as conditional variance.

Following Koening, P. (2011), in order to realize how a Power Producer is exposed to eua and fuel price co-movements, we recall the **marginal generation Cost $MC_i$**, in €/$GJ_e$ of generating a given unit of power, by using as input fuel $i$:

$$MC_i = \frac{FC_i}{n_i} + \frac{EF_i}{n_i}EC \tag{A}$$

where $FC_i$ the fuel cost in €/GJ, $n_i$ the power plant net thermal efficiency in $GJ_e$/GJ ($GJ_e$ is the power output in gigajoule of electricity, GJ the power input in gigajoule of fuel), $EF_i$ the Green House Gas (GHG) emission factor in kg $CO_2$/GJ and EC is the GHG emission cost in €/ kg $CO_2$. Equation (A) is actually a simplification and $MC_i$ is primarily estimated by the **variable costs of fuel and $CO_2$.**

The variance of $MC_i$ is given by:

$$\sigma_{MC_i}^2 = \frac{1}{n_i^2}\sigma_{FC_i}^2 + \frac{EF_i^2}{n_i^2}\sigma_{EC}^2 + 2\frac{1}{n_i}\frac{EF_i}{n_i}\rho_{FC_i,EC}\sigma_{FC_i}\sigma_{EC} \qquad (B)$$

where $\rho_{FC_i,EC}$ is the correlation of input fuels and eua and $\sigma_i^2$ are variances. Equation (B) is a **risk measure**, related to $MC_i$. In this paper will show that the pairwise correlations between electricity, fuel and eua are time-varying and also will examine how the volatility in Energy (commodity) markets in combination with volatility in financial markets affect the above conditional correlations.

## 2.2 The correlation of Carbon emission allowances (eua) with other commodity prices (ngasUK, Brent, Coal)

The optimal **merit** order of power generation is affected by changes in the relative price of input fuels. These changes ultimately result in a **fuel-switch**, by the power generator which tries to maximize its profit. Fuel-switching is not an observable operational variable and has to be inferred from changes occurred in the relative marginal costs.

From the above we conclude that the unobserved fuel-switching behavior by generators is the main factor of "producing" the correlation between input fuels (brent, ngasUK) and carbon emission allowances (eua). The empirical Carbon price moves between two extreme values, the **upper** bound **theoretical switch price $SP_u$** defined as the price of $CO_2$ above which natural gas is the preferred input fuel (technology), no matter what the thermal characteristics of the generation mix (or plant portfolio). $SP_u$ is given by

$$SP_u = \frac{n_{coal}^E FC_{gas} - n_{gas}^I FC_{coal}}{n_{gas}^I EF_{coal}^E - n_{coal}^E EF_{gas}^I} \qquad (1)$$

where $n_{coal}^E$ and $EF_{coal}^E$ are the **thermal efficiency** and **emission factor** of the most **efficient coal fired** power plant in a Country's generation mix (plant portfolio). The **thermal efficiency** and **emission factor** of the most **inefficient** gas fired power plant are $n_{gas}^I$, $EF_{gas}^I$ respectively. Therefore, **if the price of carbon increases then it will motivate generators to switch input fuels from Coal (Lignite) to gas**. As soon as $CO_2$ price has attained $SP_u$, even generators that have a choice between the most **inefficient gas** and most **efficient** Coal plat, will have, at the end, to "move" to natural gas generation. So, there is **no other** technology feasible **generation mix** which prefers coal over gas generation. An electricity producer, a profit maximizing "rational" market player, will **switch** generation from using Coal (lignite) to using natural gas, just in the case the empirical emission price exceeds the $SP_u$.

The **lower bound** theoretical switch price, $SP_l$, is the price of Carbon below which Coal is the preferred input fuel, irrespective of the thermal characteristics of the generation mix.

$$SP_l = \frac{n_{coal}^I FC_{gas} - n_{gas}^E FC_{coal}}{n_{gas}^E EF_{coal}^I - n_{coal}^I EF_{gas}^E} \qquad (2)$$

where $n_{coal}^I$, $EF_{coal}^I$ the thermal efficiency and emission factor, respectively, of the most **inefficient** coal fired plant in a Country's generation mix. $n_{gas}^E$ and $EF_{gas}^E$ are the thermal efficiency and emission factor, respectively, of the most efficient natural gas fired plant in the Country's generation mix.

Thus, if the Carbon price decreases it will give the motivation to generator to switch input fuels from natural gas to Coal power generation. When carbon price reaches $SP_l$, all generation "players" will have to switch to Coal, even though they have the choice between the most **inefficient Coal** and the most **efficient** natural gas plant.

From the above, the main conclusion is that a higher share of Coal production (Lignite in the case of Greece), rationally, will increase the demand for Carbon emission allowances (eua) and its price will upwards again.

Combining all the above the empirically observed EUA (eua time series) is expected to move between the two time-varying extreme values, $SP_l$ and $SP_u$. From the definitions given by (1) and (2), two **correlation regimes** are possible between eua and other commodities (ngasUK, Brent, Coal (API), Lignitep). The first is when eua (empirical carbon price) either exceeds $SP_u$ or falls below $SP_l$, a situation referred as **Static merit order**. In this case either natural gas or Coal are clearly the preferred input fuels and small changes in their prices **do not change the merit order.** In this case there is no financial motivation to switch input fuels, which results in an unchanged demand for eua and eua and fuel prices are **decoupled**. The **second correlation regime** is when eua is between $SP_l$ and $SP_u$.

Here we have a **mixed merit order** in which there is no clear ranking of the input fuels in the merit order and the crucial now factor in choosing one of the two fuels is their thermal efficiencies. This is a situation where **small fuel price changes have a strong influence in the merit order,** which in turn result in changes of demand for eua. This fuel and eua prices are **coupled** (or co-move). The coupling and decoupling of eua and fuel prices have been studied in depth by Koening P. (Koening, P., 2011). A very important conclusion from his work is that *if in a period t the relative forward (futures) fuel and eua prices are in such levels that make a constant merit order, then these prices are decoupled, exhibiting a low correlation.* The above situation calls for an alternative hedging strategy for securing a profit one month ahead, in comparison with a situation with **coupled** prices and **strong correlation.**

### Carbon and Commodity (Energy) markets (literature)

In theory, the equilibrium allowance price is equal to the **marginal abatement costs** incurred to reduce one ton of pollutant (Springer, 2003). The papers by Rubin (1996) and Tietenber (2006) describe the theoretical basis of deterministic equilibrium models and the solution, in a cap-and-trade framework, of the firm's pollution cost optimization problem. Thus, the participants of the market take only these measures whose costs are less than or equal to the EUA price. The theoretical justification of linking Carbon and Commodity (Energy) markets lies in the difficulty to find proxies for the emission abatement costs of a firm and their availability.

A rational abatement method is the **fuel switching** described in sections 2.1 and 2.2 (Delarue, E. et al., 2008). This method allows the power producers to abate emissions without reducing the output or making new Power plants and also take advantage of the fact that within EU ETS market the dominant player are the Power firms (representing almost 70% of the total allowances, Trotignon and Delbose, 2008). Therefore, it is expected that input fuel prices and Carbon prices must be correlated, according to the requirement of an **efficient market.**

A number of papers that provide empirical evidence about the link between Carbon and Energy (related to abatement cost) prices are shortly described below.

Commodity prices i.e. gas, oil, coal and electricity have a strong effect in the determination of Carbon prices, in Phase I of the EU ETS, as shown in the papers of Mansanet-Bataller et al. (2007), Alberola et al. (2008), and Hintermann (2010). **Positive impact** on EUA prices is shown to have **gas** and **oil prices (except Hintermann, 2010)**. However, the positive impact of Oil prices is ambiguous because this influence may be also attributed to **fuel switching** effect, to the **correlation** between **oil price** and overall **macroeconomic conditions** or to the **Oil-gas** price correlation (Rickels et al., 2010).

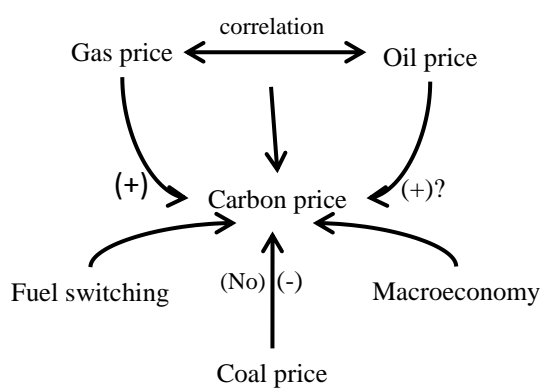

While Rickels et al. (2007) and Alberola et al. (2008) have found a negative influence of Coal price on EUA price, Mansanet-Bataller et al. (2007) and Hintermann (2010) observe no influence at all of the Coal price. The significant impact of **energy** price **volatility** on **Carbon** price **volatility**, has been found in the work of Chevallier et al. (2011), applying a univariate GARCH approach.

The influence of **regulatory interventions** also on the EUA market (particularly in Phase I) has been stressed by Alberola et al. (2008) and Hintermann (2010).

Yearly compliance events combined with Regulatory (institutional) and the macroeconomic uncertainties in EUA market have a strong influence on dynamic development of EUA price volatility (Chevallier, 2011a). In the same work, no evidence was found on the effect of the **financial crisis in 2008** on the Carbon price volatility.

For the second Phase II in the EU ETS, on which we focus in this paper, Rickels et al. (2010) confirm significant influence of gas, coal and oil prices on EUA and particularly they report a surprisingly positive impact of Coal price. Bredin and Muckley (2011), using a cointegration approach, have found an equilibrium linkage (in a new price regime in Phase II) between **Carbon futures prices** and Energy prices. The main conclusion from this work is that in Phase II the **Carbon-energy Co-movement is reinforced,** in parallel with a structural increase in correlation patterns.

Another group of literature is concentrated on the **mutual interactions** between Carbon and Energy market, considering the bi-directional influence. By using a cointegrated VAR method, Fezzi and Bunn (2009) report that **gas price** affects the **EUA price** and both jointly affect the equilibrium price of electricity in the UK market. In opposite direction are the results of Nazifi and Milunovich (2010). They instead find just short-run linkages (s.r.l) between Carbon and Oil, Carbon and gas, and electricity and Carbon, and no long-run relationship between Carbon, energy and electricity prices.

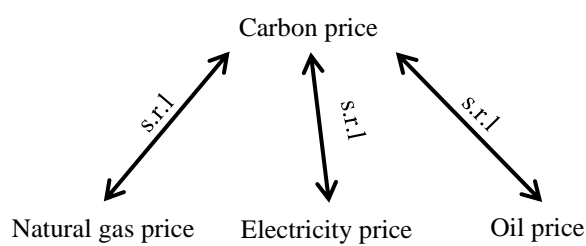

A Granger causality tests were performed by Keppler and Mansanet-Bataller (2010), and found that during Phase II, **electricity prices** Granger cause **Carbon** prices.

**Volatility spillover** between **Carbon** and **Energy** was examined by Mansanet-Bataller and Soriano (2009) using a BEKK-GARCH model. They found that **Carbon volatility** is directly and indirectly (via covariance) affected by the **Oil** and **natural gas volatility.** Carbon volatility is also affected by shocks coming from Carbon and Oil markets.

## 2.3 The interaction of financial and EUA markets

Koch, N. (2014) has found that EUA and financial markets are not isolated. Rather, financial market conditions impact strongly the correlations and the vstoxx index serves as an informative state variable reflecting the risk of "genetic" financial turmoils related to extreme events in the stock markets. According to Koch, N. (2014), the correlation between EUA Stock and Bonds (eua, ase, gbonds in our case) is expected to be strongly affected by an expected high volatility. The correlation fluctuates upwards (downwards) with peaks reverting around the collapse of Lehman Brother. He also found an impressive commonality in the EUA-Brent and EUA-Stock time-varying linkages, indicating that the positive impact of Brent Oil is possibly due to the interaction of Brent Oil prices and the overall macroeconomic situation and not due to the fuel switching (see below) or Oil-Natural gas correlation.

It is well known that macroeconomic conditions (economic growth) affect heavily both EUA and financial markets. An increased demand and raised industrial production is the result of high economic activity, which is turn increases Carbon emissions therefore increases EUA allowances (Ellerman and Buchner, 2008). Alberola et al. (2009) provide evidence of a moderate effect of Industrial production on EUA prices. Considering the Stock index as a "physical" economic indicator, Hintermann (2010) has found no significance influence of Stock Index on EUA prices in Phase I of the EU ETS, while Bonacina et al. (2009) confirm that there is a correlation between EUA (eua) prices and Euro Stoxx 50 (stoxx) in the first trading year of Phase II. Chevallier (2009) document that some particular economic factors like default spread, dividend yield or short-term interest rate are weakly correlated with EUA price, although these factors have a good forecasting power in Stock, Bond and commodity markets. Such common influences on the EUA market are not evident as Bessembinder and Chan (1992) have also observed. Furthermore, Daskalakis et al. (2009) provides strong evidence on significant negative unconditional correlations between EUA and Stock markets during 2005-2007. On the opposite, Gronwald et al. (2011) provide a strong positive Carbon-Stock markets depend-ence, which is higher for Brent Oil and Natural gas, by using Copula analysis. The impact of financial market turmoil on EUA market correlation with Stock price indices is assessed in the paper by Kanamura (2010). A multivariate correlation model was applied and provided evidence of an increased correlation in times of stock market plunge, called also **contagion**. The paper also suggests a reduction in correlation during the oversupply event, occurred in April 2006.

**Carbon and financial markets**

The Carbon market, therefore, can be characterized as a peculiar market, not influenced heavily be macroeconomic variables, and that the supply and demand of allowances is the main mechanism setting the equilibrium prices.

On the other hand, Borak et al. (2006), Benz and Truck (2006) consider Carbon as a "new" input production variable that increases the cost of generation therefore exerting pressure and uncertainties on the profits thus on the Stock market as well. They argue that EUA and Stock exhibit an indirect correlation.

The Stock market effect of the EU ETS is examined also by Veith et al. (2009) and surprisingly they identified a positive correlation between EUA prices and Stock price returns of "big" European Utilities.

The way with which the inclusion of EUAs in an assets portfolio improves the investment opportunity is examined by Mansanet-Bataller (2011) in which he finds that the opportunity set does not vary with the inclusion of Phase II EUAs, a result opposed to the one found by Chevallier, J. (2009b). It is shown, furthermore, in the above two latter papers that EUA returns are slightly negative and statistically non-significantly correlated with fixed-income securities (like Government Bonds). This result in combination with Koch, N. (2014) results is our motivation to include the Greek Government Bonds in this study, using a DCC model as opposed to the CCC models used in Mansanet-Bataller and Chevallier papers.

## 2.4 The interaction between $CO_2$ and electricity prices

Low electricity prices encourage higher electricity consumption, resulting in higher $CO_2$ emissions. Therefore the demand for allowances may increase in case electricity utilities are not in compliance with their initial allocation, a fact that in turn exerts strong pressure of the EUA markets. A further consequence is that the increase in $CO_2$ prices and generation costs may increase electricity prices creating the need for a demand adjustment, which of course implies some level of price elasticity.

Observed power and $CO_2$ prices are influenced also by fuel prices. If the prices of natural gas are increased then there is a strong incentive for generating base-load electricity by using more Coal – or Lignite fired-Plants, driving up, in turn, the demand for $CO_2$ allowances. It is worth to mention here that Coal-fired generating units emit almost twice as much $CO_2$ as natural gas generating units. If the situation just described is sustained and the supply of allowances is not adequate, $CO_2$ prices may increase at a level that result in a fuel switch i.e. natural gas, a cleaner fuel. This "cause and effect" relationship has predicted a lot of the early $CO_2$ price volatility due to the switching from Coal (Lignite) to gas. Using the **cointegration** approach, Bunn and Fezzi (2007) have analyzed the impact of EU ETS on electricity and gas prices.

## 3. The Data Sets

In this paper we consider daily data covering the period for Sep 10, 2007 to Mar 2014, a total of 2814 observations, counted after removing all non-common trading days or the sample period. The analysis period is divided into 2 periods: a) the **Subprime Crisis period (January 2, 2008 until the end of 2009)** and b) the **Greek Government Debt Crisis (early 2010 until April 4, 2011)**. The two periods correspond to the two shaded areas in the DCC plots (section 5.2). The chosen sampling frequency produce sufficient number of data required to measure the dynamics of correlations which may vary due to periods of financial turmoil of differing durations. The price data are denominated in the local currency of each market. To enhance our choice of data frequency, we point out that from an EU ETS participant point of view, caring for his risk management, high frequency (here daily) correlations are more useful that long-term correlations. The data sets are obtained from various resources, ASE, IPTO, ICE Futures Europe, EIA.

### 3.1 The Carbon Market and the EUA data

The three phases of the EU ETS, corresponding to the three compliance periods are Phase I: 2005-2007, Phase II: 2008-2012 and Phase III: 2013-2020. The pilot period of the EU ETS (European Union Emissions Trading Scheme) is the well-known to market participants Phase I. The National Allocation Plans (NAPs) determine the overall emission cap for phases I and Phase II. Each member state determines its NAP, defining actually the total permits and the allocation mode. NAPs are approved by European Commission (EC), which settles the overall cap. Because neither borrowing nor banking of EUA (EU Allowances) were allowed between Phase I and Phase II, the price for EUAs (series eua in this paper) issued for Phase I collapsed. The first information regarding the actual EUAs released in April 2006, however the market participants considered that the total emission cap for Phase I was not restrictive. Phases II and III are linked by banking, where the transactions of spare EUAs enlarges the time period considered by the agents when they shape their expectations about the overall shortage of EUAs. The Banking involvement reduces, therefore, the risk of an extreme collapse of the EUA price. But, if shocks happen they still can generate strong price and volatility fluctuations. Highly efficient EUA spot and derivative markets have evolved since 2005 and the most liquid derivative market is the European Climate Exchange (ICE/ECX, London), where 90% of the futures contracts are traded.

**Description of the data**

Historical Futures Prices: ECX EUA Futures, Continuous Contract #1. Non-adjusted price based on spot-month continuous contract calculations. Raw data from ICE.

Daily settlement prices of EUA futures contracts (€/ton) traded on the ICE ECX are used to form a continuous price time series that combines a number of contracts expiring in Phase II and III (2008-2012 and 2013-2020), following the approach of Koch, N. (2011). We mention here that trading of EUA futures contracts started not until April 22, 2005. The price of the 2008 contract constitutes the continuous carbon price time series during Phase I. This series changes to the December 2009 contract in Phase II, up to the last trading day, on which day the series changes again into the next yearly contract. According to Koch (2011) this method of constructing the continuous EUA series is unlikely to introduce a bias because the used futures contracts are not redeemable in Phase I. This choice in forming the EUA series in further enhanced by the fact that EUA are required only once a year, for the reason of compliance, so holding spot EUAs does not offer any advantage in comparison with holding a corresponding futures position (Daskalakis et al., 2009). Also, Koch (2011) concludes that the EUA futures prices for Phase II can be considered as the reliable "real" price signal for investors. We have used EUA data obtained from ICE ECX market because this is the leading exchange (Mizrach and Otsubo, 2011).

### 3.2 The Commodity (energy) data. Natural Gas prices at NBP hubs and the Greek Natural Gas "market"

We use daily spot price of Brent Oil traded in Euro/barrel, obtained from Bloomberg. For natural gas, daily spot prices traded at the National Balancing Point NBP Hub UK, expressed in €/MWh, is considered obtained also from Bloomberg. Since the late 1990s, UK NBP Hub gas market is Europe's longest established wholesale (spot-traded) market in operation. Natural Gas UK NBP price (ngasUK) is extensively used as the Europe's benchmark price indicator. This wholesale gas market is the most liquid one in Europe nowadays, alongside a number of newly established Continental Europe hubs (e.g. Zeebrugge in Belgium and TTF in Netherlands) NBP is the acromion for National Balancing Point and gas anywhere in UK within the NGNTS (Natural Gas National Transmission System) counts as NBP gas. This Hub brings together buyers and sellers so the trading is greatly simplified. There is a variety of products: within-day (for same day delivery), day-ahead (for next day delivery), months, quarters, summers (April to September) and winters (October to March), as well as annual contracts.

Normally, spot contracts at NBP Hub are in pence sterling per therm. In this paper we convert the prices of all the time series to Euro per megawatt (€/MWh), the standard in Europe, allowing us for a better understanding of co-variations of prices. The appropriate conversion is 1 therm per 0.0293 MWh ICIS[1], and the conversion of pence sterling to Euro is according to the daily exchange rate published by the ECB (European Central Bank)[2].

The data covers the period from Sep 2007 until Mar 2014 on a daily basis, generating 2371 observations, provided by Thomson Reuters, available on Datastream with TRGBNBD and TRBEZED tickers (mnemonics) for NBP and Zee-brugge date respectively. The data are actually indices reflecting the volume weighted average price of all transactions by the OTC trading platform, at a given day.

There is no indigenous gas production in Greece and also there are no storage facilities (the LNG storage tanks are used exclusively for temporary LNG storage, the three entry points of natural gas to the **National Natural Gas System (NNGS)** of Greece are located at Sidirocastro, Greek Bulgarian pipeline, for the Russian gas, at Kipi, Greek-Turkish pipeline (BOTAS gas) and at the Revithoussa LNG terminal. In Greece, the gas market is till organized on the basis of bilateral contracts between suppliers and eligible customers, so there is not any wholesale market yet. The Regulator RAE, published for the first time in 2011, the **Weighted-Average Import Price (WAIP)** of natural gas, on a monthly basis. This data on WAIP, considered together with the publication of data on daily prices of **balancing gas (HTAE)** on the Natural Gas TSO's (DESFA) internet site, has greatly facilitate current and potential market participants in understanding the prevailing gas price dynamics. The figure 1-a below shows the monthly average SMP, WAIP against the daily HTAE price for the same month (the daily HTAE price is kept constant over the entire month considered). Data are published on RAE's website[3] and updated on a regular basis.

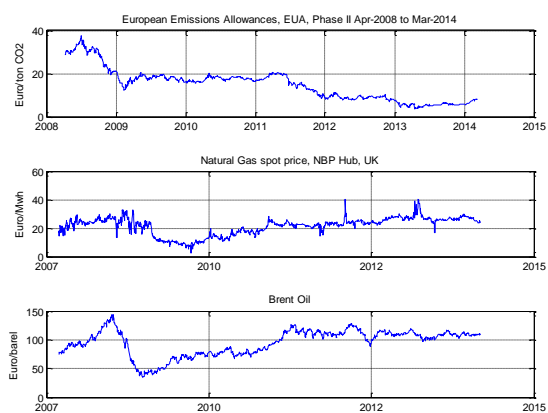

**Figure 1**: EUA, Natural gas and Brent Oil , price time series.

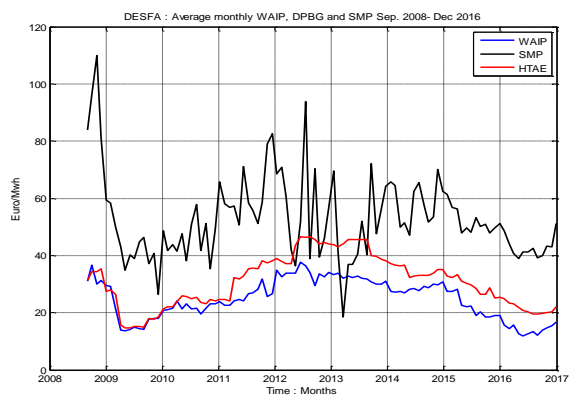

**Figure 1A**: The monthly Weighted Average Import Prices of Natural gas, the daily prices of balancing gas (HTAE) and SMP in the GEM.

---

### 3.3 The Greek Wholesale or System Marginal Price

*Greece's liberalized electricity market* was established according to the European Directive 96/92/EC and consists of two separate markets: 1) the Wholesale Energy and Ancillary Services Market and 2) the Capacity Assurance Market. The Greek wholesale electricity market (GEM) is currently in a transitional period, during which the market structure evolves towards its final design, namely the European Target Model. The wholesale electricity market is a day ahead mandatory pool which is subject to inter-zonal transmission constraints, unit technical constraints, reserve requirements, the interconnection Net Transfer Capacities (NTCs) and in general all system constraints. More specifically, based on forecasted demand, generators' offers, suppliers' bids, power stations' availabilities, unpriced or must-run production (e.g., hydro power mandatory generation, cogeneration and RES outputs), schedules for interconnection as well as a number of transmission system's and power station's technical constraints, an optimization process is followed in order to dispatch the power plant with the lower cost, both for energy and ancillary services.

LAGIE (the independent market operator) (www.lagie.gr) is responsible for the solution of the so-called Day Ahead (optimization) problem. This problem is formulated as a security constrained unit commitment problem, and its solution is considered to be the optimum state of the system at which the social welfare is maximized for all 24 h of the next day simultaneously. This is possible through matching the energy to be absorbed with the energy injected into the system, i.e., matching supply and demand (according to each unit's separate offers). The DA solution, therefore, determines the way of operation of each unit for each hour (dispatch period) of the dispatch day as well as the clearing price of the DA market's components (energy and reserves).

More specifically in this pool, market "agents" participating in the Energy component of the day-ahead (DA) market submit offers (bids) on a daily basis. Producers and importers submit energy offers with the limitation that the weighted average of the offer should be above the unit Minimum Average Variable Cost. On the contrary exporters and load representatives submit load declarations. The bids are in the form of a 10-step stepwise monotonically increasing (decreasing) function of pairs of prices (€/MWh) and quantities (MWh) for each of the 24 h period of the next day. A single price and quantity pair for each category of reserve energy (primary, secondary and tertiary) is also submitted by generators. Deadline for offer submission is at 12.00 pm ("gate" closure time).

So, the DAS solution produces a 24 hour unit schedule and a unique price which is called the System's Marginal Price (SMP). The Dispatch Scheduling (DS) is used to define the time period between Day Ahead Schedule (DAS) and Real Time Dispatch (RTD) where the producers have the chance to change their declarations whenever has been a problem regarding the availability of their units. In the RTD the units are re-dispatched in real time in order to meet the actual demand. Finally in the IS stage an Ex Post Imbalance Pricing (EXPIP) is produced after the dispatch day which is based on the actual demand and unit availability. The capacity assurance market is a procedure where each load representative is assigned a capacity adequacy obligation and each producer issues capacity availability tickets for its net capacity. Actually this mechanism is facing any adequacies in capacity and is in place for the partial recovery of capital costs. The most expensive unit dispatched determines the uniform pricing in the day-ahead market. In case of congestion problems and as a motive for driving new capacity investment, zonal pricing is a solution, but at the moment this approach has not been activated. Physical delivery transactions are bounded within the pool although market agents may be entering into bilateral financial contracts that are not currently in existence. The offers of the generators are capped by an upper price level of 150€/MWh. Physical Transmission Rights (PTR) are explicitly allocated via auctions.

Not only the fundamentals but also the various Regulatory Market Reforms (RMRs), "imposed" by the Greek Regulatory for Energy (RAE), have a significant impact on the volatilities of energy and electric prices (RAE, 2009 to 2014), Kalantzis et al., 2012, Papaioannou et al., 2017). The reforms took place on specific dates – milestones or Reference Days. We describe here only the reforms made within the period of our analysis in this paper:

**4th Reference Day (1.5.2008) (RMR5).** **Cost Recovery Mechanism, CRM,** was considered by the Regulator a necessary step until the **Imbalance Settlement Mechanism, ISM** (scheduled for the 5th Reference Day). CRM states that if the SMP is lower than the marginal cost of generating Unit (plus 10%), then the Unit will receive the difference as a compensation. The Regulator expected that this Reform would have no effect on SMP. CRM was aiming to ensure that generators will be compensated at least their marginal cost, in case they were ordered to operate. The Cost Recovery Mechanism was abolished on 30th June 2014.

**RMR6. Regulatory Market Reform, RMR6** (RAE's Decision 1.1.2009), focused on the change of the ex-post SMP calculation methodology according to the unit commitment algorithm that considers all technical constraints of the units and the reserve requirements of the IPTO (ADMIE) expecting to lead to lower SMPs.

**5th Reference Day (30.9.2010) (RMR7). Regulatory Market Reform, RMR7,** initiated the mandatory day-ahead market model and introduced the Imbalances Settlement Mechanism retaining at the same time the SMP methodology allowing the submission of demand declarations. **RMR7** is referred to the adoption of an enhanced Unit commitment algorithm which co-optimizes energy as well as ancillary services. In this new mandatory, Day-Ahead market model incorporating, at the same time, an Imbalance settlement mechanism4, market clearance is now based on the non-priced demand declarations. Taking into account that the methodology for estimating SMP retained the same and the fact that usually the declared demands were underestimated, the effect of this reform expected to reduce SMP slightly.

**RMR8. Regulatory Market Reform, RMR8** (Ministry of Finance Decision 1.9.2011), regards the decision of the Ministry of Finance (1.9.2011) to impose a new tax levy on natural gas, equal to 1.50€/GJ (applied also to electricity generation). As SMP was set, for the majority of trading periods, by Natural Gas fired Units, the resulted increased generation cost was expected to increase SMP (see section 6.1 for comments).

**RMR9. Regulatory Market Reform, RMR9** (1.7.2013), Abolition of the "Plus 10% Rule". This rule was embedded in Cost Recovery Mechanism (CRM) and allowed for a 10% increase of the boundary for generators to be compensated for generating costs.

**RMR10. Regulatory Market Reform, RMR10** (31.12.2013), Abolition of the "30% Rule". The "30% Rule" allows generators to offer 30% of their plant's capacity at a price below its minimum variable cost, as long as the total weighted average of their bids is still at or above their minimum variable cost. This caused the extended dispatch of gas plants, pushing the expenses on cost-recovery significantly high. The regulator expected no changes on the SMP through this reform, it was imposed merely to improve the performance of the initial market design.

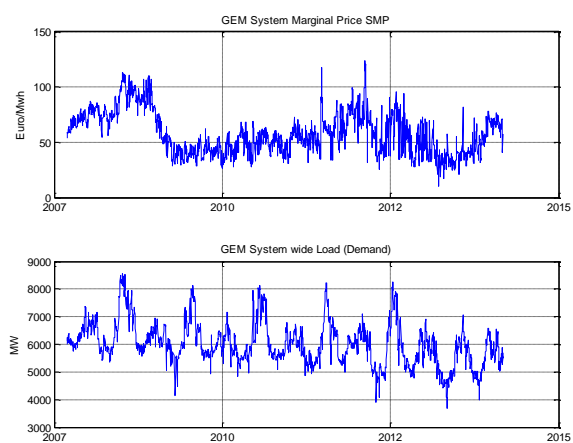

**Figure 2:** The wholesale (Day-Ahead, System Marginal Price, smp) and the system-wide load (demand) in the Greek electricity market (GEM) 10-Sep-2007 to 07-Mar-2014.

## 3.4 The Financial data

---

We have used the **Athens Stock Exchange General Index (ase),** denominated in Euro. In order to "capture" the in-dependence-interaction of the Greek Stock market with the European financial market, especially during the financial crisis period (focusing on the European sovereign debt crisis in 2010), we have considered also the EURO STOXX 50 price index, in Euro, which covers 50 blue-chip stock from 12 euro are countries (Austria, Belgium, Finland, France, Germany, Greece, Ireland, Italy, Luxembourg, the Netherlands, Portugal and Spain). We choose this particular index, following Koch (2011), because it is the basis for the EURO STOXX 50 Volatility Index (vstoxx), reflecting the market expectations of volatility. It measures the square root of implied variance over all EURO STOXX 50 options, for the next 30 days. Measuring the so-call investor's fear in case that it is larger than 30 indicates a large amount of volatility, reflecting the investor's uncertainty or fear.

For bond, we use the 10-year Greek Government bond index (gbonds) (a long-term index), instead of a short-term index, because monetary policy (especially during the Greek financial Crisis) is move likely to have a confounding impact on the later index.

We include also in the financial data set the stock price of the dominant player in GEM, the incubator Public Power Corporation (PPC). We consider that by analyzing the dynamic evolution of this stock we "capture" the various effects of regulatory policy and fundamental changes, exerted by monetary (macroeconomic) policies to fix the Greek Public Debt problem as well as European Energy Policies.

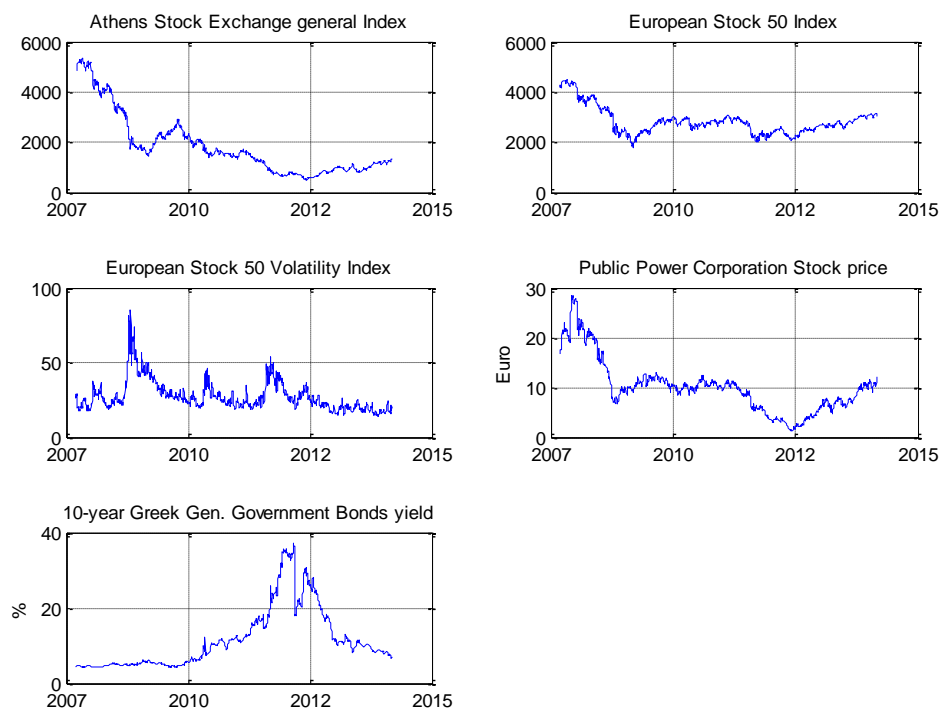

**Figure 3:** The financial data set's "assets", price (levels) time series for the period 10-Sep-2007 to 07-Mar-2014.

**Table 2:** The Data Sets containing the variables employed in the DDC Analysis (10-Sep-2007 to 07-Mar-2014)

| | Name | Description |
|---|------|-------------|
| **Financial Data Set** | | |
| 1 | ase: | Athens Stock Exchange General Index |
| 2 | stoxx: | European Stock 50 Index (Euro stoxx 50) |
| 3 | vstoxx: | European Stock 50 Volatility Index (vstoxx) |
| 4 | ppc: | Public Power Corporation (PPC) Stock price |
| 5 | gbonds: | Greek Government 10 year Bond yield |
| **Energy Commodities Data Set** | | |
| 1 | eua: | European Union Allowance (EU Emissions Trading Scheme): €/tCO2 |
| 2 | ngas: | Natural gas price, NBP, UK, and Zeebrugge Hub, Belgium €/MWh |
| 3 | brent: | Brent Oil price, $/bbl |
| **Power (Electricity) Data Set** | | |
| 1 | smp: | Greek Electricity Market wholesale or System Marginal Price (ex-ante) (€/MWh) |
| 2 | load: | Electricity load (Mw) (ex-post) |
| 3 | Lignite price | Lignite Operational Cost of a "typical" Lignite-fired Power plant (€/MWh) in North Greece |

# 4. Financial and Econometric methodology

## Using a VAR modelling the Conditional mean

The equation or model of the **Conditional mean** or **first moment** is to detect and eliminate any serial correlation in the returns of price data. As we have seen in section, by applying the Ljung-Box test statistics, there is strong evidence of significant serial correlation in the returns. Vector Autoregression (VAR) of lag order p is used in this paper to estimate the first moment.

Let $\mathbf{r}_t$ symbolizes a kx1 vector of returns at time t, $\mathbf{r}_t=\{r_{i,t}\}$, where $r_{i,t}$ is the daily log returns, for i=1….k. The VAR(p) model is written as

$$\mathbf{r}_t=\mathbf{\Phi}_0+\sum_{j=1}^{p}\mathbf{\Phi}_j\mathbf{r}_{t-j}+\mathbf{\varepsilon}_t \qquad \text{or}$$

$$\mathbf{r}_t=\mathbf{\Phi}_0+\mathbf{\Phi}_1\mathbf{r}_{t-1}+\ldots+\mathbf{\Phi}_p\mathbf{r}_{t-p}+\mathbf{\varepsilon}_t \qquad (A)$$

where $\mathbf{\Phi}_0$ is a kx1 vector of constants, $\mathbf{\Phi}_j$ kxk matrix of coefficients and $\mathbf{\varepsilon}_t$ a kx1 vector of residuals. The "optimum" lag length p of the VAR(p) can be found by minimizing the **Akaike Information Criterion (AIC).** The specification then of the "best" model, based on AIC, is accepted if the residual "pass" successfully a number of diagnostic tests (e.g. checking for remaining serial correlation).

As an example, let k=3, a **trivariate** model $\mathbf{r}=(\boldsymbol{ase}, \boldsymbol{stoxx}, \boldsymbol{vstoxx})'$ and let p=2, lags, then (A) becomes

$$\begin{bmatrix} ase,t \\ stoxx,t \\ vstoxx,t \end{bmatrix} = \begin{bmatrix} \Phi_{1,0} \\ \Phi_{2,0} \\ \Phi_{3,0} \end{bmatrix} + \begin{bmatrix} \Phi_{11,1} & \Phi_{12,1} & \Phi_{13,1} \\ \Phi_{21,1} & \Phi_{22,1} & \Phi_{23,1} \\ \Phi_{31,1} & \Phi_{32,1} & \Phi_{33,1} \end{bmatrix} \cdot \begin{bmatrix} ase,t-1 \\ stoxx,t-1 \\ vstoxx,t-1 \end{bmatrix} + \begin{bmatrix} \Phi_{11,2} & \Phi_{12,2} & \Phi_{13,2} \\ \Phi_{21,2} & \Phi_{22,2} & \Phi_{23,2} \\ \Phi_{31,2} & \Phi_{32,2} & \Phi_{33,2} \end{bmatrix} \cdot \begin{bmatrix} ase,t-2 \\ stoxx,t-2 \\ vstoxx,t-2 \end{bmatrix} + \begin{bmatrix} \varepsilon_{ase,t} \\ \varepsilon_{stoxx,t} \\ \varepsilon_{vstoxx,t} \end{bmatrix}$$

Serially uncorrelated residuals are generated by a well-specified model for the first moment of the returns. However, heteroskedasticity (the time-varying variance of the residuals) will remain in the returns, as it is frequently the case in Energy and financial markets. This feature and the excess kurtosis in the returns call for the GARCH-type estimation approach (Engle, 1982, Bollerslev, 1986). The GARCH model incorporates the heteroskedasticity characteristic of the data. The works of Chevallier et al. (2009), Benz and Trueck (2009), Mansanet-Bataller and Soriano (2009) refer to the application of this type of model in Carbon (EUA) and energy market time series.

Let that the mean of a return time series follows an autoregressive of order p, AR(P), specification

$$r_{i,t} = a_o + \sum_{j=1}^{P} a_j r_{i,t-j} + \varepsilon_{i,t} \tag{1}$$

where $r_{i,t}$ is the daily log returns of $K$ time series for $i=1,...,K$, $\varepsilon_{i,t}$ is the **residual** of series $i$ and $a_o$ the drift term. Suppose that $F_{t-1}$ is the set of all available information about the process, up to the time $t-1$, then the **conditional variance** of the residual $\varepsilon_{i,t}$ is $\sigma_{i,t}^2$, so $\varepsilon_{i,t}|F_{t-1} \sim N(0, \sigma_{i,t}^2)$ or $\varepsilon_{i,t} = \sigma_{i,t} n_t$ where $n_t \sim NID(0,1)$.

This $\varepsilon_{i,t}$ residual is fitted in the GARCH-type models, described below, to capture the dynamics of the **conditional variance.**

Let the evolution of the **conditional variance** in the generic univariate process for each asset, is written as

$$\sigma_\delta^2 = \omega + \sum_{p=1}^{P} \alpha_p |\varepsilon_{t-p}|^\delta + \sum_{o=1}^{O} \gamma_o |\varepsilon_{t-o}|^\delta I[\varepsilon_{t-o} < o] + \sum_{q=1}^{Q} \beta_q \sigma_{t-q}^\delta \tag{2}$$

where $\delta$ is either *1* for threshold ARCH also known as AVGARCH, ZARCH (Taylor, 1986, Zakoian, 1994) or *2* for ARCH, GARCH or GJR-GARCH models (Glosten et al., 1993). In this paper we consider the case of $\delta=2$ and particularly the case GJR-GARCH(P,O,Q). In fact, we fit our data in a GJR-GARCH(1,1,1) model, the dynamics of which is written as

$$\sigma_t^2 = \omega + \alpha_1 \varepsilon_{t-1}^2 + \gamma_1 \varepsilon_{t-1}^2 I_{[\varepsilon_{t-1} < 0]} + \beta_1 \sigma_{t-1}^2 \tag{3}$$

where $I_{[\varepsilon_{t-1} < 0]}$ is an indicator function that takes the value 1 if $\varepsilon_{t-1} < 0$ and $\boldsymbol{0}$ otherwise. This function takes care of the **asymmetries** of the impact on volatility the returns may have due to "good" or "bad" news. The parameters must be such that $\omega > 0$, $\alpha_1 \geq 0$, $\alpha_1 + \gamma \geq 0$ and $\beta_1 \geq 0$, and for the covariance to be stationary, $\alpha_1 + \frac{1}{2}\gamma_1 + \beta_1 < 1$ (**mean reverting** model). In case $\alpha_1 + \beta_1 = 1$ we have an **integrated** model.

In estimating $h_{it}$ from univariate volatility models, the BIC Schwartz Information Criterion is use to select suitable candidate models that capture the stylized facts of the asset return.

## 4.1 Constant Conditional Correlation (CCC) and Dynamic Conditional Correlation, DCC, Models

A multivariate GARCH(P,O,Q) is a natural extension of the univariate model, and allows for the time-varying correlations between two series, in addition to their conditional variances. To generate a vector of residuals (hopefully serially uncorrelated) we could use a Vector Autoregression model, VAR(p), to model the mean of a 10X1 vector consisting of the members of the financial, energy and power group of data set, given in Table 2. The model produces the following vector of residuals

$$\boldsymbol{\varepsilon}_t = \left( \varepsilon_{ase,t}, \varepsilon_{stoxx,t}, \varepsilon_{vstoxx,t}, \varepsilon_{ppc,t}, \varepsilon_{gbonds,t}, \varepsilon_{eua,t}, \varepsilon_{ngas,t}, \varepsilon_{brent,t}, \varepsilon_{smp,t}, \varepsilon_{load,t} \right)'$$

We also suppose that the underlying distribution of returns follows a conditional multivariate normal process, therefore we can write $\varepsilon_t | F_{t-1} \sim N(\boldsymbol{0}, H_t)$, where $F_{t-1}$ is a filtration i.e. an information set about the time series up to the time step $t-1$. Thus, the $\varepsilon_t$ is conditionally heteroskedastic, which means that $\varepsilon_t = \sqrt{H_t} \cdot \boldsymbol{n}_t$, where $\boldsymbol{n}_t \sim N(\boldsymbol{0}, I)$ an iid error process.

For modelling $H_t$ a number of specifications has been suggested, the most commonly mentioned is the generic VECH-model, developed by Bollerslev et al. (1988), the **CCC-model (Constant Conditional Correlation)** also by Bollerslev (1990) and the BEKK-model by Engle and Kroner (1995). A detailed survey on multivariate GARCH models is provided by Silvennoinen and Tersvirta (2007).

In this paper will apply the parsimonious **Dynamic Conditional Correlation (DCC)** approach, developed by Engle (2000) and Engle and Sheppard (2001). This model is actually a natural extension of the CCC-model, giving the opportunity for a two-stage estimation of the dynamic evolution of conditional correlations between, for example, two commodities. In the first stage of the procedure, standardized residuals are generated by univariate GARCH models fitted on the data of the individual time series. In the second stage the correlation process is estimated.

According to the work of Engle and Sheppard (2001), the **conditional covariance matrix** $H_t$ is written as follows

$$H_t = D_t R_t D_t \tag{4}$$

where $D_t$ a kXk diagonal matrix with elements $\sqrt{\sigma_{i,t}^2}$ on the $i$th diagonal representing the time-varying standard deviations which are generated by the GARCH models fitted on each residual series, as the ones given in equation (2). $R_t$ is the time-varying **conditional correlation matrix.** In the case of **CCC-model** we have:

$$\textbf{\textit{Model 1:}} \ \ H_t = D_t R D_t \tag{5}$$

$$R = (\boldsymbol{\rho_{ij}})$$

where $R$=Constant Conditional Correlation. The assumption that conditional correlations are constant is unrealistic in particular applications, although the estimation of CCC parameters is more simple. We use CCC hare as a benchmark for testing the consistency of correlations (see Table 8 below).

The log-likelihood is our case, for the vector $\theta$ of parameters is given by

$$L(\theta) = -\frac{1}{2} \sum_{t=1}^{T} \left( m \log(2\pi) + 2 \log(|D_t|) + \log(|R_t|) + \xi_t' R_t^{-1} \xi_t \right) \tag{6}$$

where $\xi_t \sim N(\boldsymbol{0}, R_t)$ the standardized residuals, $\xi_t = \frac{\varepsilon_t}{D_t}$.

In case that the conditional distribution of $\varepsilon_t$ is not normal, equation (4) is the Quasi-likelihood function. The dynamic correlation specification suggested by Engle and Sheppard (2001) is:

$$Q_t = \left( I - \sum_{j=1}^{P} \alpha_j - \sum_{j=1}^{Q} \beta_j \right) \bar{Q} + \sum_{j=1}^{P} \alpha_j (\xi_{t-j} \ddot{\xi}_{t-j}) + \sum_{j=1}^{Q} \beta_j Q_{t-j} \tag{7}$$

where $\bar{Q}$ is the kXk **unconditional covariance matrix** of the standardized residuals, generated from the first stage of the process. The extent to which $\xi_t$ affect the dynamics of the correlation is captured by the $\alpha_j$, while $\beta_j$ is a parameter measuring the **decay** in dynamics. If we plug $\alpha_j = \beta_j = 0$ into (6), the CCC model of Bollerslev (1990) is obtained. The lag-lengths of residuals and decay are expressed in $P$ and $Q$ (not to be confused with those in equation (1)). Finally, the dynamic conditional correlation is written

$$R_t = Q_t^{*-1} Q_t Q_t^{*-1} \tag{8}$$

where $Q_t^*$ is a diagonal matrix (kXk) consisting of the square root of the diagonal elements of $Q_t$. Furthermore, the conditional covariance matrix $R_t$ of the residuals generated by VAR(p), is obtained by standardizing these residuals by the conditional variances, so a typical element of $R_t$ is

$$\rho_{i,j,t} = \frac{q_{i,j,t}}{\sqrt{q_{i,i,t}q_{j,j,t}}} \qquad (9)$$

In the framework of this paper estimation, the indices range as

$$i,j = ase, \ stoxx, vstoxx, ppc, eua, ngas, brent, smp, load, \textbf{\textit{lignitep}}$$

By letting $P=Q=1$ in equation (6) we obtained our **DCC model 2** specification:

$$\textbf{\textit{Model 2}} : Q_t = (1-\alpha-\beta)\overline{Q} + \alpha\left(\xi_{t-1}\xi_{t-1}^{'}\right) + \beta Q_{t-1} \qquad (10)$$

The matrix $Q_t$ is a symmetric positive matrix, $\alpha+\beta<1$, and $\alpha$ is the **news coefficient** and $\beta$ is the **decay** coefficient. According to Aielli (2011), typical values of the dynamic parameters $\alpha, \alpha+\beta$ are $\alpha+\beta>0.80$ and $\alpha\leq0.04$ while in financial application, in particular, $\alpha+\beta\geq0.96$ and $\alpha\leq0.04$. $\overline{Q}=E\left[\xi_t\xi_t^{'}\right]$ is the unconditional correlation (the unconditional variance matrix of the standardize residuals. A typical element of the correlation matrix $R_t$, regarding the interaction, for example, between ase index and ppc stock price is

$$\rho_{ase,ppc,t} = \frac{q_{ase,ppc,t}}{\sqrt{q_{ase,ppc,t}q_{ase,ppc,t}}}$$

Therefore, by using model 2 above, we have

$$q_{ase,ppc,t} = (1-\alpha-\beta)\overline{q}_{ase,ppc} + \alpha\left(\xi_{ase,t-1}\xi_{ppc,t-1}^{'}\right) + \beta q_{ase,ppc,t-1}$$

$$q_{ase,t} = (1-\alpha-\beta)\overline{q}_{ase} + \alpha\left(\xi_{ase,t-1}^2\right) + \beta q_{ase,t-1}$$

$$q_{ppc,t} = (1-\alpha-\beta)\overline{q}_{ppc} + \alpha\left(\xi_{ppc,t-1}^2\right) + \beta q_{ppc,t-1}$$

Model 1 will be our basic reference model. This scalar DCC specification is the most parsimonious one because of the assumption that all commodities correlations "obey" the same ARMA(P,Q) type specification, which means that they are all governed by the same coefficients $\alpha$ and $\beta$. The above assumption might be a valid one, in the case of similar commodities (or "assets" in general), belonging in same asset category or class. However, in our case, our "assets" belong to different categories, namely financial, energy and power; therefore it is a reasonable assumption that these markets exhibit "asset" specific correlation sensitivities. To face this dissimilarity in asset's class, a generalization of the DCC model has been suggested, incorporating also the impact of any asymmetries on the correlation dynamics. It is known that in a Markov Switching Model (MSM) or in a Threshold Autoregressive Model (TARM), the conditional correlations are allowed to have different evolutionary dynamics. Instead this is not the case for DCC model in which the correlations follow the same dynamics. This is a limitation of the DCC. For example, if the data exhibit structural breaks, DCC model can give misleading conclusions. Another limitation of DCC is that it does not work reliably for large number of assets. Cappielo et al. (2006) have developed a number of various asymmetric multivariate GARCH models to capture the asymmetries. For an in depth description of the "mathematical" properties, its limitation and inconsistencies in DCC model, Aielli (2011) provides an excellent work.

## 4.2 The Asymmetric Generalized DCC Model

[Engle (2009)] propose a **Generalized Dynamic Conditional Correlation (G-DCC)** in order to tackle the correlation across asset categories, a flexible model allowing for asset specific correlation parameters. The model is written as

$$\textit{Model 3:} \quad Q_t = \left(\bar{Q} - A'\bar{Q}A - B'\bar{Q}B\right) + A'\xi_{t-1}\xi'_{t-1}A + B'Q_{t-1}B \tag{11}$$

where $A$ and $B$ are kXk diagonal matrices of the parameters, $A = \{\alpha_{ii}\}$, $B = \{\beta_{ii}\}$.

The positive definiteness requirement is satisfied by $\alpha_{ii} + \beta_{ii} < 1$ and $\alpha_{ii}, \beta_{ii} \geq 0$, $\forall i,j$. The above specification tackles the dissimilarity of asset problem by allowing for a high degree of dissimilarity in correlations.

The advantage of G-DCC over the simple scalar DCC is that it can generate a variety of correlation patterns. The coefficients $\alpha_{ii}$ can be considered are measuring the sensitivity of the correlation of asset $i$ with other assets to correlation residuals [Hafner and Frances, 2003]. High values for $\alpha_{ii}$ in combination with low values for $\beta_{ii}$ result in almost horizontal, very flat correlations of asset $i$ with any other asset in the specification. Instead, low values for $\alpha_{ii}$ combined with high values for $\beta_{ii}$ produce very fluctuating correlations.

[Cappielo et al. (2006)] propose a further generalization, the **AG-DCC (Asymmetric Generalized DCC)** model 4 that actually nests model 4, written as

$$\textit{Model 4:} \quad Q_t = \left(\bar{Q} - A'\bar{Q}A - B'\bar{Q}B - G'\bar{N}G\right) + A'\xi_{t-1}\xi'_{t-1}A + B'Q_{t-1}B + G'\boldsymbol{n}_{t-1}\boldsymbol{n}'_{t-1}G \tag{12}$$

where $G$ is a kXk diagonal matrix of parameters, $G = \{g_{ii}\}$, $\boldsymbol{n}_t = \{n_{i,t}\}$ a kX1 vector with $n_{i,t} = min(\xi_t, 0)$, $\bar{N}$ is a kXk matrix of constants, $\bar{N} = T^{-1}\sum_{t=1}^{T}\boldsymbol{n}_t\boldsymbol{n}'_t$.

Similarly as in model 3, the positive definiteness requirement is satisfied by $\alpha_{ii} + \beta_{ii} + n_i k < 1$ and $\alpha_{ii}, \beta_{ii}, n_i \geq 0$, $for\ i=1,....,k$ where $k$ is the maximum eigenvalue of $\sqrt{\bar{Q}}\bar{N}\sqrt{\bar{Q}}$ [Cappielo et al., 2006].

Model 4 is further extended to include control ("exogenous") variables [Vargas (2008)] proposed the AG-DCC-X model and it is this model used by [Koening (2011)] to test the hypothesis of the effect of static merit order regimes on correlation between input fuels, carbon emission and electricity prices. We do not consider the model in this paper. By using that $A'^*A = A^2$, $B'^*B = B^2$ etc., little algebra transforms model 3 into the following form

$$Q_t = \left(1 - A^2 - B^2\right)\bar{Q} + A^2\xi_{t-1}\xi'_{t-1} + B^2 Q_{t-1} + G^2\left(\boldsymbol{n}_{t-1}\boldsymbol{n}'_{t-1} - \bar{N}\right) \tag{13}$$

# 5. Empirical findings

## 5.1

**Table 3:** Summary Statistics of Price (levels) (10-Sep-2007 to 07-Mar-2014)

| Price (level) Series | | ase | stoxx | vstoxx | ppc | gbonds | eua | ngas | brent | smp | load | lignitep |
|---|---|---|---|---|---|---|---|---|---|---|---|---|
| Observations | | 2371 | | 2371 | 2371 | 2371 | 2371 | 2371 | 2371 | 2371 | 2371 | 2160 |
| Mean | | 1844.38 | 2827.71 | 27.14 | 10.03 | 11.50 | 13.60 | 21.46 | 94.91 | 57.51 | 6027.93 | 33.69 |
| Median | | 1516.95 | 2744.18 | 24.37 | 9.82 | 9.27 | 14.95 | 23.07 | 102.68 | 54.94 | 5961.96 | 35.10 |
| Maximum | | 5334.50 | 4489.79 | 85.44 | 28.44 | 37.10 | 37.43 | 40.29 | 143.95 | 123.77 | 8555.83 | 48.53 |
| Minimum | | 476.36 | 1809.98 | 13.82 | 1.15 | 4.22 | 1.03 | 2.64 | 34.45 | 10.24 | 3684.54 | 21.98 |
| Std. Dev. | | 1215.24 | 535.51 | 10.08 | 5.41 | 8.03 | 8.07 | 6.10 | 22.67 | 19.11 | 781.08 | 5.77 |
| Skewness | | 1.35 | 1.30 | 1.93 | 1.12 | 1.41 | 0.47 | -0.60 | -0.66 | 0.56 | 0.55 | -0.20 |
| Kurtosis | | 3.98 | 4.63 | 8.01 | 4.39 | 4.23 | 2.88 | 2.85 | 2.68 | 2.77 | 3.48 | 2.76 |
| JB (p-value) | h | 1.00 | 1.00 | 1.00 | 1.00 | 1.00 | 1.00 | 1.00 | 1.00 | 1.00 | 1.00 | 1.00 |
| | p-value | 0.00 | 0.00 | 0.00 | 0.00 | 0.00 | 0.00 | 0.00 | 0.00 | 0.00 | 0.00 | 0.00 |
| | Stat. | 812.33 | 929.95 | 3953.83 | 683.52 | 939.46 | 89.40 | 142.64 | 180.72 | 127.79 | 141.80 | 19.27 |
| ADF | h | 1.00 | 0.00 | 0.00 | 0.00 | 0.00 | 0.00 | 0.00 | 0.00 | 1.00 | 0.00 | 0.00 |
| | p-value | 0.00 | 0.27 | 0.11 | 0.27 | 0.37 | 0.31 | 0.20 | 0.68 | 0.00 | 0.33 | 0.84 |
| | Stat. | -3.22 | -1.04 | -1.57 | -1.03 | -0.77 | -0.92 | -1.23 | -0.09 | -3.20 | -0.89 | 0.59 |
| PP | h | 1.00 | 0.00 | 0.00 | 0.00 | 0.00 | 0.00 | 0.00 | 0.00 | 1.00 | 0.00 | 0.00 |
| | p-value | 0.00 | 0.27 | 0.11 | 0.27 | 0.37 | 0.31 | 0.20 | 0.68 | 0.00 | 0.33 | 0.84 |
| | Stat. | -3.22 | -1.04 | -1.57 | -1.03 | -0.77 | -0.92 | -1.23 | 0.09 | -3.20 | -0.89 | 0.59 |

**Table 3a:**

| | 'ase' | 'stoxx' | 'vstoxx' | 'ppc' | 'gbonds' | 'eua' | 'ngUK' | 'brent' | 'lignite' | 'smp' | 'loadep' |
|---|---|---|---|---|---|---|---|---|---|---|---|
| 'ase' | 1.0000 | 0.6504 | 0.0851 | 0.8790 | -0.6797 | 0.8527 | -0.3235 | -0.2276 | -0.7683 | 0.2562 | 0.3134 |
| 'stoxx' | | 1.0000 | -0.5302 | 0.7323 | -0.4838 | 0.4718 | 0.1816 | 0.3467 | -0.1803 | 0.1619 | 0.0633 |
| 'vstoxx' | | | 1.0000 | -0.0913 | -0.0307 | 0.2897 | -0.2652 | -0.5587 | -0.4267 | 0.3028 | 0.1319 |
| 'ppc' | | | | 1.0000 | -0.7478 | 0.7086 | -0.2704 | -0.1821 | -0.4764 | 0.1143 | 0.2468 |
| 'gbonds' | | | | | 1.0000 | -0.4271 | 0.3159 | 0.4740 | 0.3777 | 0.1455 | -0.0716 |
| 'eua' | | | | | | 1.0000 | -0.2455 | -0.1550 | -0.8159 | 0.4379 | 0.4122 |
| 'ngUK' | | | | | | | 1.0000 | 0.6536 | 0.3461 | 0.2587 | -0.1552 |
| 'brent' | | | | | | | | 1.0000 | 0.3749 | 0.1706 | -0.0773 |
| 'lignite' | | | | | | | | | 1.0000 | -0.3414 | -0.3944 |
| 'smp' | | | | | | | | | | 1.0000 | 0.4675 |
| 'loadep' | | | | | | | | | | | 1.0000 |

**Table 4:** Daily log returns Summary Statistics (10-Sep-2007 to 07-Mar-2014).

| Log Return Series | | ase | stoxx | vstoxx | ppc | gbonds | eua | ngas | brent | smp | load | lignitep |
|---|---|---|---|---|---|---|---|---|---|---|---|---|
| **Panel A: Descriptive statistics** | | | | | | | | | | | | |
| Observations | | 2371 | 2371 | 2371 | 2371 | 2371 | 2160 | 2371 | 2371 | 2371 | 2371 | 2159 |
| Mean | | 0.000 | -0.000 | -0.00 | 0.00 | 0.00 | -0.00 | 0.00 | 0.00 | 0.00 | -0.00 | 0.00 |
| Median | | 0.000 | 0.000 | 0.00 | 0.00 | 0.00 | 0.00 | 0.00 | 0.00 | 0.00 | 0.00 | 0.00 |
| Maximum | | 1.0180 | 0.10 | 0.33 | 0.22 | 0.14 | 0.19 | 0.73 | 0.18 | 1.02 | 0.20 | 0.29 |
| Minimum | | -0.8673 | -0.08 | -0.27 | -0.25 | -0.68 | -0.34 | -1.06 | -0.17 | -0.87 | -0.25 | -0.25 |
| Std. Dev. | | 0.1551 | 0.01 | 0.05 | 0.03 | 0.02 | 0.02 | 0.07 | 0.02 | 0.16 | 0.04 | 0.02 |
| Skewness | | 0.0233 | 0.08 | 0.85 | -0.09 | -12.96 | -1.10 | -1.39 | 0.03 | 0.02 | -0.63 | 4.21 |
| Kurtosis | | 9.4841 | 11.4 | 7.95 | 10.64 | 352.9 | 29.06 | 49.05 | 16.16 | 9.48 | 10.41 | 103.69 |
| JB (p-value) | h | 1 | 1 | 1 | 1 | 1 | 1 | 1 | 1 | 1 | 1 | 1.00 |
| | p-value | 0.00 | 0.00 | 0.00 | 0.00 | 0.00 | 0.00 | 0.00 | 0.00 | 0.00 | 0.00 | 0.00 |
| | Stat. | 4152 | 6992 | 2700 | 5774 | $12.15 \times 10^7$ | 61548 | 210145 | 17092 | 4152 | 5583 | 918408.54 |
| **Panel B: Stationarity** | | | | | | | | | | | | |
| ADF | h1 | 1 | 1 | 1 | 1 | 1 | 1 | 1 | 1 | 1 | 1 | 0.00 |
| | p-value | 0.000 | 0.000 | 0.000 | 0.00 | 0.00 | 0.00 | 0.00 | 0.00 | 0.00 | 0.00 | -58.80 |
| | Stat. | -65.85 | -48.70 | -48.91 | -45.7 | -44.14 | -44.4 | -59.8 | -49.15 | -65.8 | -54.6 | 1.00 |
| PP | h | 1 | 1 | 1 | 1 | 1 | 1 | 1 | 1 | 1 | 1 | 0.00 |
| | p-value | 0.000 | 0.000 | 0.000 | 0.000 | 0.000 | 0.000 | 0.000 | 0.00 | 0.00 | 0.00 | -58.80 |
| | Stat. | -65.85 | -48.70 | -48.91 | -45.7 | -44.14 | -44.4 | -59.8 | -49.15 | -65.8 | -54.6 | 1.00 |
| **Panel C: Serial Correlation. ARCH tests** | | | | | | | | | | | | |
| Q(20) | h | 1 | 1 | 1 | 1 | 1 | 1 | 1 | 1 | 1 | 1 | 187.97 |
| | p-value | 0.000 | 0.000 | 0.000 | 0.000 | 0.000 | 0.000 | 0.000 | 0.00 | 0.00 | 0.00 | 1.00 |
| | Stat. | 265.22 | 52.65 | 54.27 | 41.52 | 63.14 | 58.05 | 188.5 | 42.62 | 265 | 183.9 | 0.00 |
| $Q^2$(20) | h | 1 | 1 | 1 | 1 | 0 | 1 | 1 | 1 | 1 | 1 | 187.97 |
| | p-value | 0.000 | 0.000 | 0.000 | 0.000 | 1.000 | 0.000 | 0.000 | 0.00 | 0.00 | 0.00 | 1.00 |
| | Stat. | 465.5 | 885.22 | 164.8 | 282.7 | 1.36 | 133.6 | 483 | 418 | 465.5 | 137.4 | 0.00 |
| ARCH-LM(20) | h | 1 | 1 | 1 | 1 | 0 | 1 | 1 | 1 | 1 | 1 | 247.30 |
| | p-value | 0.000 | 0.000 | 0.000 | 0.000 | 1.000 | 0.000 | 0.000 | 0.00 | 0.00 | 0.00 | 1.00 |
| | Stat. | 31.41 | 420.65 | 124.5 | 186.32 | 1.30 | 31.4 | 414 | 226 | 277 | 132.30 | 0.00 |

Q(20) and Q²(20) are Ljung-Box or Q statistics for detecting the existence of autocorrelation in returns and squared returns respectively for the first 20 lays. The 5% critical values of X²(20) distributions is 31.41. For ADF and PP test, the 1% critical value is -3.44

By observing table 4 we conclude that financial, energy and electricity "asset" returns are likely to be non-Gaussian. In all returns the skewness is non-zero, an evidence of a non-symmetric distribution. Farther more, the kurtosis is significantly in excess (>3) which indicates fat tails of the distribution, containing more probability than a normal distribution. In combination, the log returns of "assets" are **leptokurtic**. Also, we test the return for normality by applying the Jarque-Bera (JB) test statistic. According to this test the joint null hypothesis is that both skewness and excess kurtosis are zero. As we observe in table 4, the p-value for the JB statistics is zero in all returns, therefore the null hypothesis can safely be rejected, so the returns follow a non-normal distribution.

We have also applied the **Augmented-Dickey-Fuller (ADF)** and **Phillips-Perron (PP)** unit root tests. As we observe, both tests give values larger than the critical value for the 1% level of significance. Therefore, we can reject the null-hypothesis of a unit root for all returns, so they are taken to be stationary.

To detect autocorrelations in the returns we have used the Ljung-Box or Q statistic. From table 4 also, we see that all returns show signs of statistically significant autocorrelation, and only the Greek Government Bond (gbonds) show the opposite (the Q(20) and Q²(20) statistics are less than 31.41, p-value=1000). The strongest autocorrelation is in ase, smp, load and ngas returns (with Q(20) statistics 265.22, 265, 183.9 and 188.5 respectively).
All log returns have a mean zero. GEM wholesale price (smp) returns are most volatile (std. Dev $\simeq$ 0.16) followed by ase and ngas. The least vitality is stoxx, with other assets having comparable volatilities.

All returns show evidence of volatility clustering (ARCH effects) as the visual inspection of the log return (see fig. 4-6) and the **ARCH-Lagrangean Multiplier (ARCH-LM)** test in table 4, show. With an exception of gbonds returns, the test statistic is significant higher than its critical value for the 5%. So, statistically significant ARCH effects exist in those returns. gbonds returns, however, show no formal sign of ARCH effects.

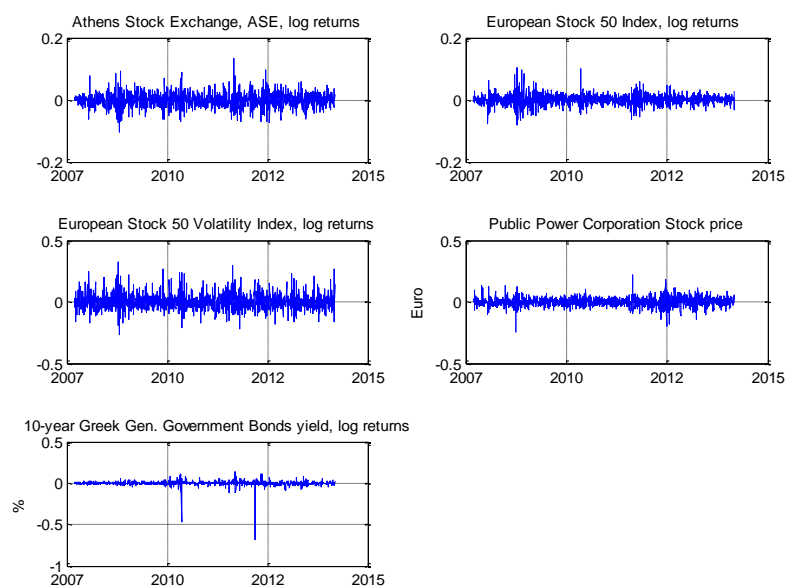

**Figure 4**: log-returns of the financial set's time series for the period 10-Sep-2007 to 07-Mar-2014.

The figures 4-6 depict the dynamic evolution of the returns of the time series used. All returns are characterized by the well-known phenomena of **volatility clusters**. Furthermore, as the figures show, during the aftermath of the **Lehman Brothers bankruptcy**, September 2008, and during the **European Sovereign Debt Crisis**, mid 2010, all returns exhibit high level of volatility and the associated clustering. The sample autocorrelation function of the squared returns (not shown here due to space limitation) is slowly decaying, a typical feature for daily returns exhibiting volatility clustering. The ARCH-LM test results, mentioned before, confirm the existence of this stylized fact.

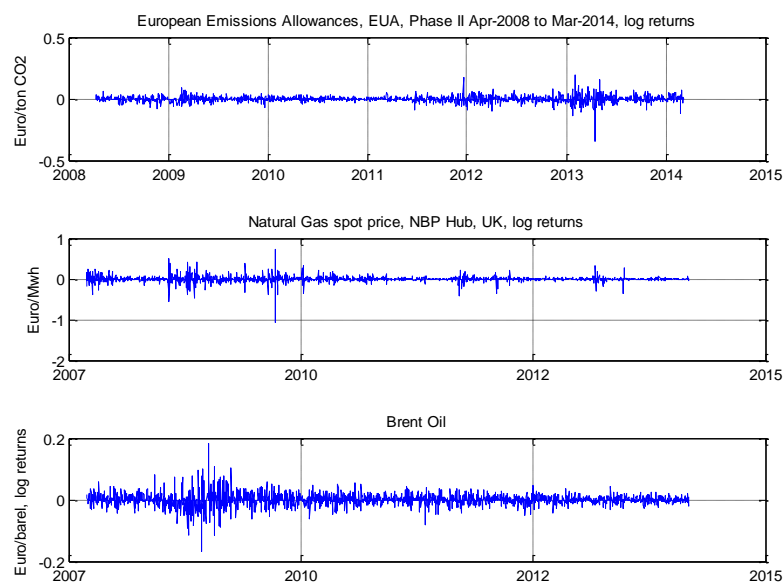

**Figure 5:** Log-returns of the Energy (commodity) data set's time series for the period 10-Sep-2007 to 07-Mar-2014.

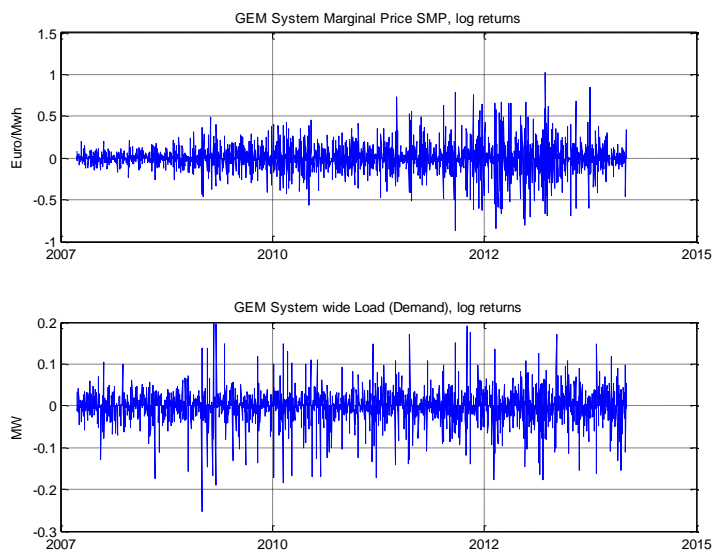

**Figure 6:** Log-returns of the Electricity data set's time series for the period 10-Sep-2007 to 07-Mar-2014.

**Table 5:** Estimated parameters in the GARCH(1,1) specification.

| GARCH (1.1) Parameters | | | | |
|---|---|---|---|---|
| **Residuals** | **ω** | **α₁** | **β₁** | **α₁+β₁** |
| ase_res | 0.0000 | 0.0428 | 0.9382 | 0.9810 |
| stoxx_res | 0.0000 | 0.0602 | 0.9258 | 0.9860 |
| ppc_res | 0.0000 | 0.0467 | 0.9360 | 0.9827 |
| vstoxx_res | 0.0010 | 0.0436 | 0.9071 | 0.9507 |
| gbonds_res | 0.0000 | 0.0846 | 0.9152 | 0.9998 |
| eua_res | 0.0000 | 0.0930 | 0.9068 | 0.9998 |
| ngasUK_res | 0.0000 | 0.1110 | 0.8887 | 0.9998 |
| brent_res | 0.0000 | 0.0312 | 0.9671 | 0.9983 |
| smp_res | 0.0000 | 0.0204 | 0.9791 | 0.9995 |
| loadep_res | 0.0050 | 0.0963 | 0.5069 | 0.6035 |
| lignitep_res | 0.0000 | 0.0317 | 0.8415 | 0.8732 |

**Table 6**: Log-likelihood (LL) for the estimated GARCH specifications

| **Residuals** | **GARCH (1.1)** | **GJR-GARCH(1.1)** | **Mean Equation Specification** | | | |
|---|---|---|---|---|---|---|
| | | | **AR** | **MA** | **AIC** | **LL** |
| ase_res | 6233.13 | 6246.63 | 0 | 1 | 6118 | 6135 |
| stoxx_res | 7113.16 | 7187.93 | 4 | 7 | 6775 | 6800 |
| ppc_res | 5183.00 | 5206.53 | 7 | 7 | 4992 | 5022 |
| vstoxx_res | 3768.00 | 3793.64 | 5 | 5 | 3676 | 3704 |
| gbonds_res | 5692.26 | 5875.50 | 6 | 2 | 5501 | 5523 |
| eua_res | 2882.85 | 2913.58 | 2 | 2 | 4614 | 4632 |
| ngasUK_res | 3946.95 | 3955.21 | 5 | 7 | 3170 | 3197 |
| brent_res | 6607.29 | 6621.79 | 7 | 7 | 6171 | 6201 |
| smp_res | 1591.78 | 1608.96 | 4 | 3 | 1261 | 1283 |
| loadep_res | 4699.73 | 4716.40 | 7 | 7 | 4614 | 4644 |
| lignitep_res | | | | | | |

**Table 7**: CCC and DCC GARCH(1,1) estimation results

| Pair of "Assets" residuals | | Estimated Parameters | | | | |
|---|---|---|---|---|---|---|
| | | CCC | DCC(1,1) | | | |
| | | Model 1 | Model 2 equation (10) | | | |
| | | ρ | α | β | α+β | LL |
| ase_res | stoxx_res | 0.4584 | 0.0000 | 0.9999 | 0.9999 | 12393 |
| ase_res | vstoxx_res | -0.3150 | 0.0000 | 0.9995 | 0.9995 | 9171 |
| ase_res | ppc_res | 0.6283 | 0.0061 | 0.9933 | 0.9994 | 10886 |
| ase_res | gbonds_res | -0.2030 | 0.0381 | 0.9557 | 0.9939 | 10798 |
| ase_res | ngasUK_res | 0.0297 | 0.0000 | 0.0000 | 0.0000 | 9225 |
| ase_res | brent_res | 0.1791 | 0.0125 | 0.9790 | 0.9915 | 11666 |
| ase_res | eua_res | 0.0936 | 0.0076 | 0.9809 | 0.9885 | 10719 |
| ase_res | lignite_res | 0.0336 | 0.0056 | 0.9524 | 0.9580 | 11456 |
| gbonds_res | vstoxx_res | 0.1201 | 0.0154 | 0.9814 | 0.9967 | 8532 |
| eua_res | ngasUK_res | 0.0657 | 0.0216 | 0.7605 | 0.7820 | 8696 |
| eua_res | brent_res | 0.2200 | 0.0144 | 0.9738 | 0.9882 | 11146 |
| eua_res | ppc_res | 0.0367 | 0.0121 | 0.9291 | 0.9412 | 9767 |
| eua_res | vstoxx_res | -0.1507 | 0.0148 | 0.9607 | 0.9755 | 8540 |
| eua_res | smp_res | 0.0077 | 0.0362 | 0.3200 | 0.3562 | 6291 |
| eua_res | loadep_res | 0.0343 | 0.0091 | 0.0002 | 0.0094 | 9213 |
| smp_res | ngasUK_res | 0.0281 | 0.0000 | 0.8788 | 0.8788 | 4813 |
| smp_res | loadep_res | 0.2771 | 0.0139 | 0.9304 | 0.9443 | 5426 |
| loadep_res | ase_res | -0.0699 | 0.0294 | 0.2705 | 0.2999 | 9751 |
| smp_res | ase_res | -0.0211 | 0.0074 | 0.9190 | 0.9264 | 6823 |
| lignitep_res | smp_res | -0.0560 | 0.0186 | 0.9236 | 0.9422 | 7049 |
| lignitep_res | eua_res | -0.0151 | 0.0032 | 0.9913 | 0.9946 | 10921 |
| lignitep_res | ngasUK_res | -0.0113 | 0.0000 | 0.0001 | 0.0001 | 9445 |
| lignitep_res | brent_res | 0.0085 | 0.0133 | 0.4322 | 0.4455 | 11830 |
| lignitep_res | ppc_res | 0.0177 | 0.0000 | 0.0004 | 0.0004 | 10519 |
| gbonds_res | brent_res | -0.0041 | 0.0092 | 0.9861 | 0.9952 | 11073 |

Table 7 shows the estimation for the DCC-GARCH(1,1) model. The coefficients α and β for all pairs of "assets" are positive as required to ensure positive unconditional variances. Also, in all cases, α+β<1 (except for the pair ase_res and gbonds_res). We have mentioned already that if α+β~1 then the DCC is **highly persistence.** So, pairs of assets having α+β close to 1, like ase_res and stoxx_res, ase_res and ppc_res, etc., exhibit a high persistent DCC i.e. they strongly co-move or there is a strong volatility spillover between the two assets. The pairs with the lowest persistence in their DCC are ase_ngasUK, eua_smp and lignite_brent, with α+β equal to 0.25. 0.45, 0.54, respectively.

Table 7 shows also, the CCC-model's parameter ρ, with positive and negative values. The strongest positive constant conditional correlation is for ase_ppc pair, reflecting a strong linkage between the fluctuations in Athens Stock Exchange (ASE) general Index and the Public Power Corporation's, PPC, Stock price (ppc). Strong and positive (0.47) between ase and stoxx is estimated, while the volatility index vstoxx and ase are negatively correlated. This negative correlation is also shown in Figure 19 depicting the DCC between also those two "assets". An explanation for this is given below in the results section.

## 5.2 CCC and DCC between Power and Energy commodities

**Table 8:** Summary Statistics of DCC estimates

| Pair of assets | Total Sample | | | Financial Crisis Sample | | | Greek Government Debt Crisis | | |
|---|---|---|---|---|---|---|---|---|---|
| | **Mean** | **Min** | **Max** | **Mean** | **Min** | **Max** | **Mean** | **Min** | **Max** |
| 'ase_stoxx' | 0.448 | 0.215 | 0.717 | 0.630 | 0.547 | 0.717 | 0.486 | 0.434 | 0.539 |
| 'ase_vstoxx' | -0.313 | -0.537 | -0.173 | -0.448 | -0.537 | -0.372 | -0.325 | -0.365 | -0.289 |
| 'ase_ppc' | 0.618 | 0.097 | 0.834 | 0.480 | 0.097 | 0.602 | 0.564 | 0.499 | 0.646 |
| 'ase_gbonds' | -0.268 | -0.735 | 0.485 | -0.054 | -0.632 | 0.485 | -0.423 | -0.731 | -0.019 |
| 'ase_ngUK' | 0.028 | 0.028 | 0.028 | 0.028 | 0.028 | 0.028 | 0.028 | 0.028 | 0.028 |
| 'ase_brent' | 0.184 | -0.256 | 0.432 | 0.194 | -0.256 | 0.409 | 0.255 | 0.060 | 0.432 |
| 'ase_eua' | 0.090 | -0.498 | 0.280 | 0.101 | -0.498 | 0.280 | 0.091 | 0.004 | 0.159 |
| 'ase_lignite' | 0.037 | -0.077 | 0.085 | 0.033 | -0.077 | 0.069 | 0.042 | -0.003 | 0.079 |
| 'gbonds_vstoxx' | 0.135 | -0.520 | 0.520 | -0.053 | -0.520 | 0.112 | 0.169 | 0.013 | 0.386 |
| 'eua_ngUK' | 0.064 | -0.198 | 0.309 | 0.058 | -0.147 | 0.265 | 0.072 | -0.041 | 0.309 |
| 'eua_brent' | 0.210 | -0.069 | 0.502 | 0.302 | 0.118 | 0.502 | 0.136 | 0.005 | 0.349 |
| 'eua_ppc' | 0.042 | -0.136 | 0.255 | 0.061 | -0.064 | 0.255 | 0.037 | -0.056 | 0.155 |
| 'eua_vstoxx' | -0.145 | -0.361 | 0.198 | -0.156 | -0.336 | 0.198 | -0.119 | -0.285 | 0.149 |
| 'eua_smp' | 0.015 | -0.367 | 0.276 | 0.016 | -0.172 | 0.260 | 0.015 | -0.185 | 0.180 |
| 'eua_loadep' | 0.034 | -0.070 | 0.123 | 0.033 | -0.054 | 0.084 | 0.033 | -0.070 | 0.108 |
| 'smp_ngUK' | 0.026 | 0.002 | 0.027 | 0.026 | 0.002 | 0.027 | 0.027 | 0.027 | 0.027 |
| 'smp_loadep' | 0.280 | 0.003 | 0.512 | 0.294 | 0.067 | 0.512 | 0.283 | 0.174 | 0.391 |
| 'loadep_ase' | -0.068 | -0.389 | 0.292 | -0.067 | -0.278 | 0.103 | -0.066 | -0.263 | 0.094 |
| 'smp_ase' | -0.021 | -0.108 | 0.082 | -0.016 | -0.066 | 0.082 | -0.017 | -0.086 | 0.044 |
| 'lignite_smp' | -0.070 | -0.242 | 0.125 | -0.078 | -0.242 | 0.025 | -0.063 | -0.210 | 0.066 |
| 'lignite_eua' | -0.014 | -0.236 | 0.046 | -0.035 | -0.236 | 0.010 | -0.029 | -0.067 | 0.000 |
| 'lignite_ngUK' | -0.011 | -0.011 | -0.011 | -0.011 | -0.011 | -0.011 | -0.011 | -0.011 | -0.011 |
| 'lignite_brent' | 0.007 | -0.095 | 0.104 | 0.007 | -0.043 | 0.104 | 0.006 | -0.095 | 0.103 |
| 'lignite_ppc' | 0.018 | 0.018 | 0.018 | 0.018 | 0.018 | 0.018 | 0.018 | 0.018 | 0.018 |
| gbonds_brent | -0.061 | -0.269 | 0.322 | 0.044 | -0.085 | 0.322 | -0.148 | -0.268 | -0.046 |

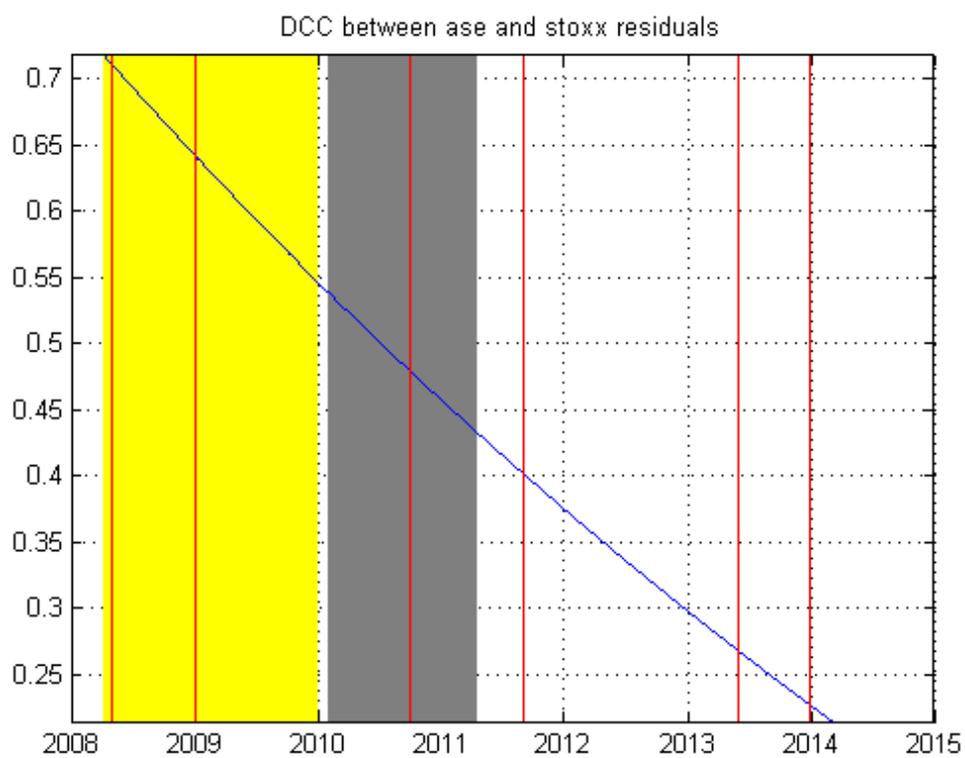

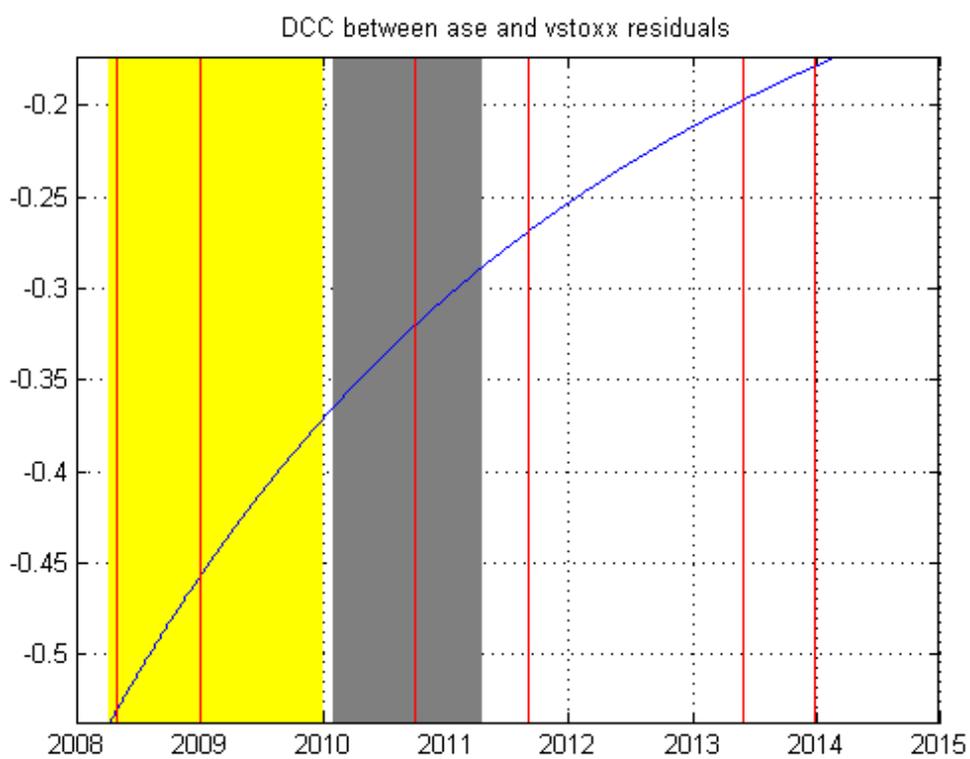

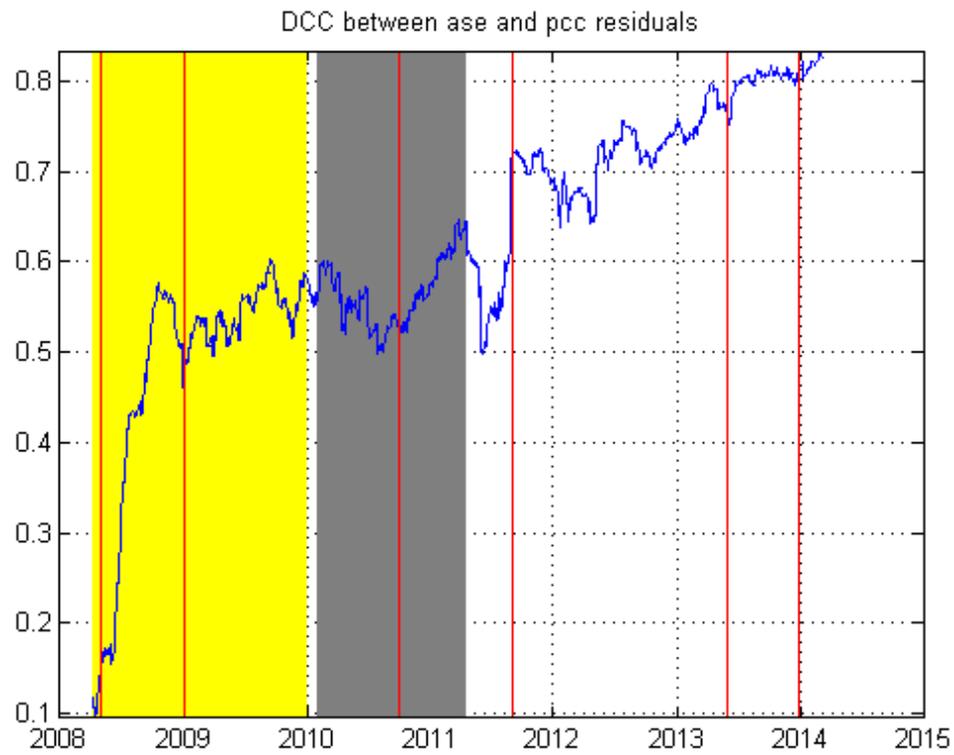

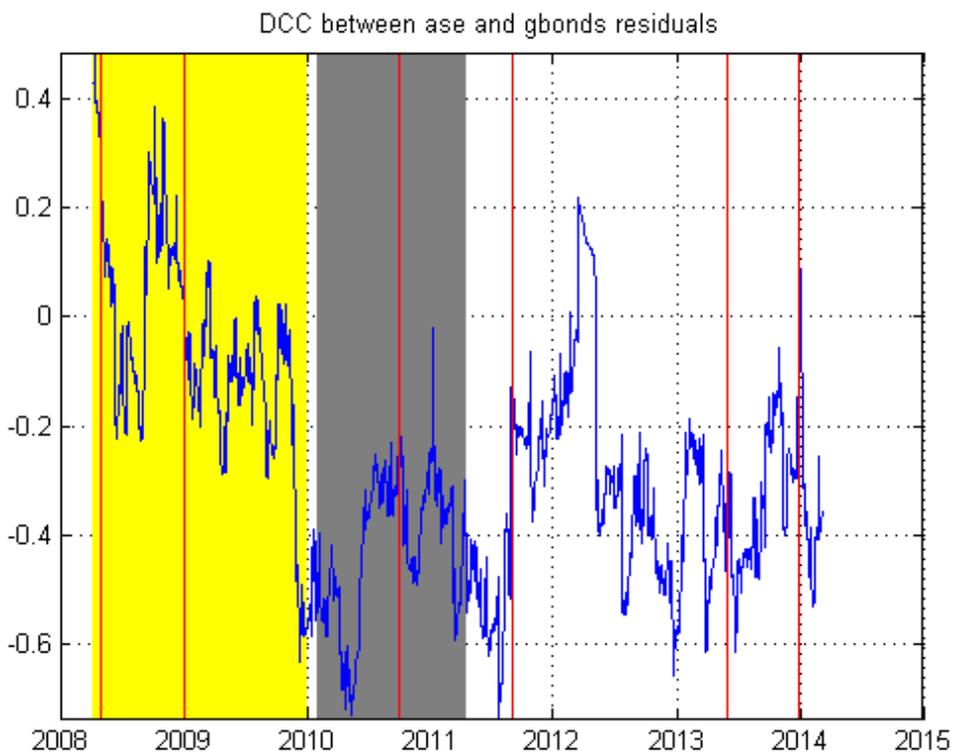

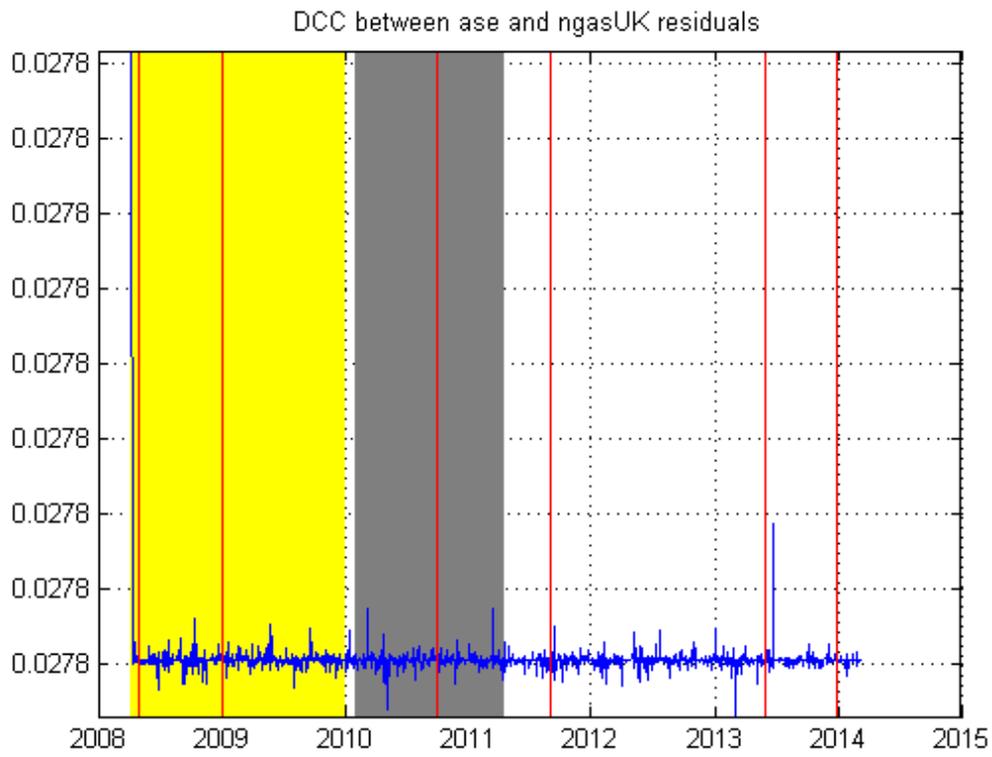

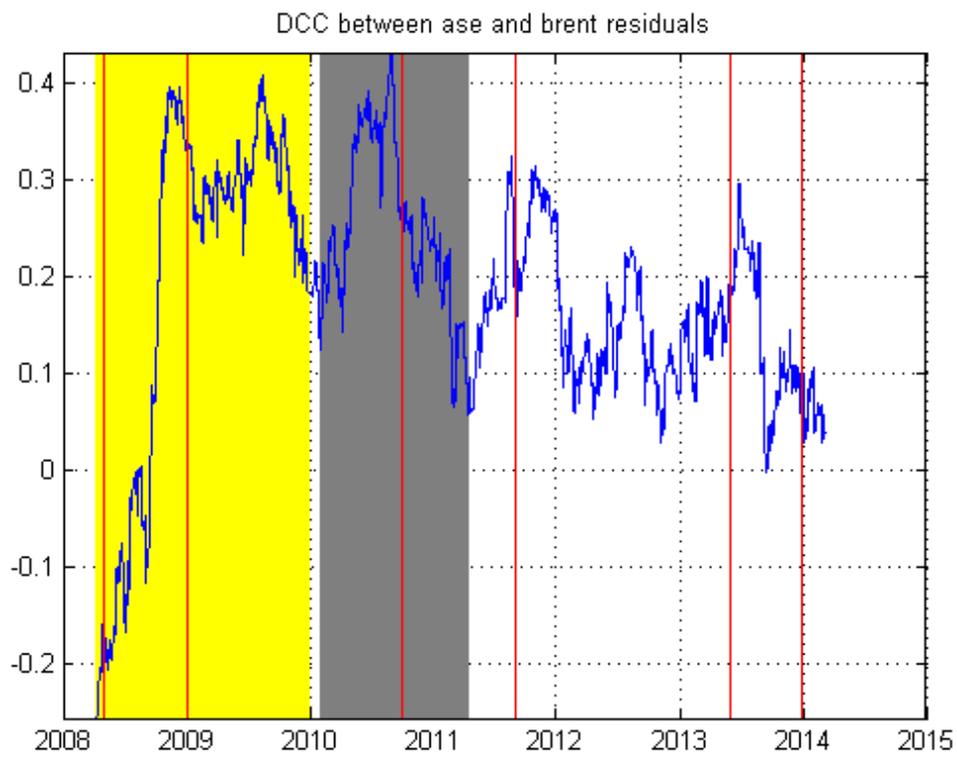

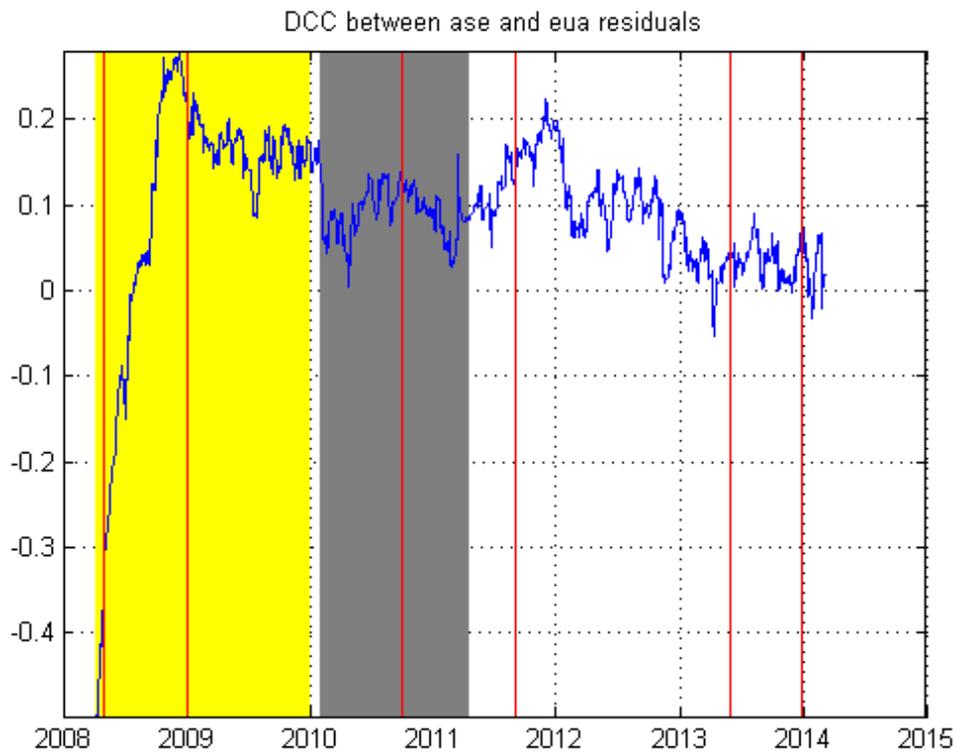

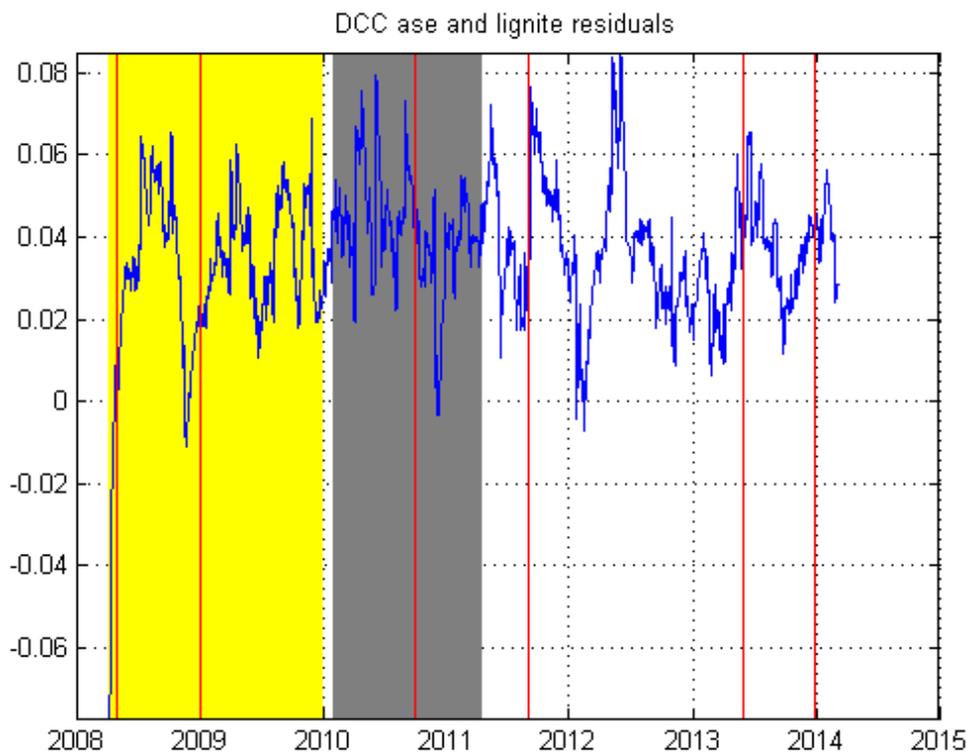

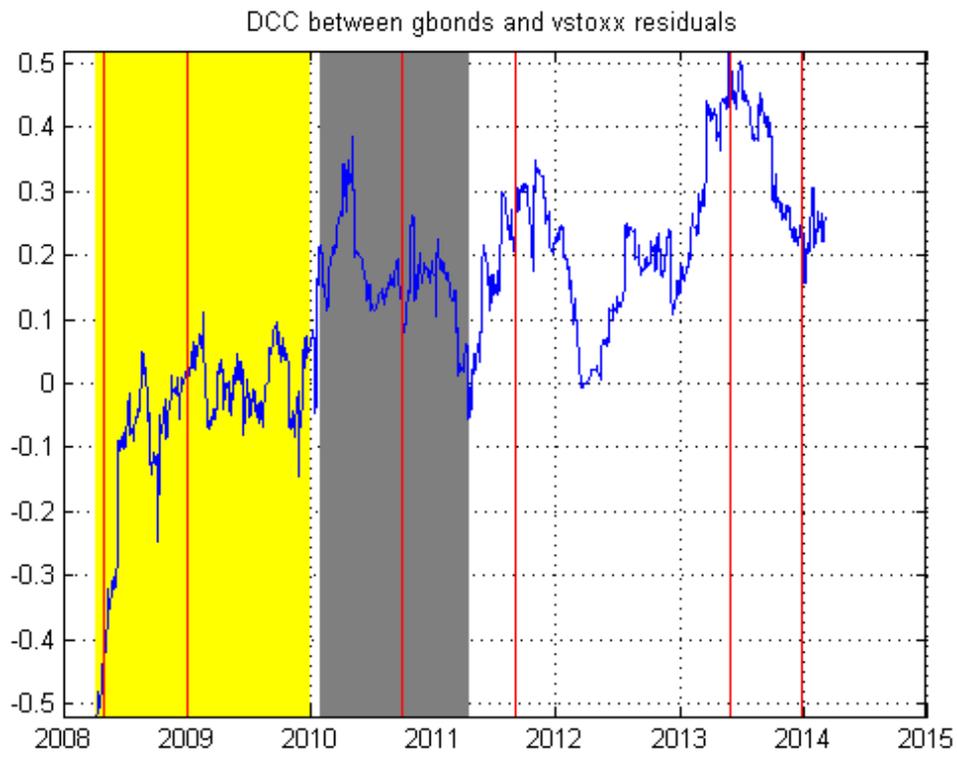

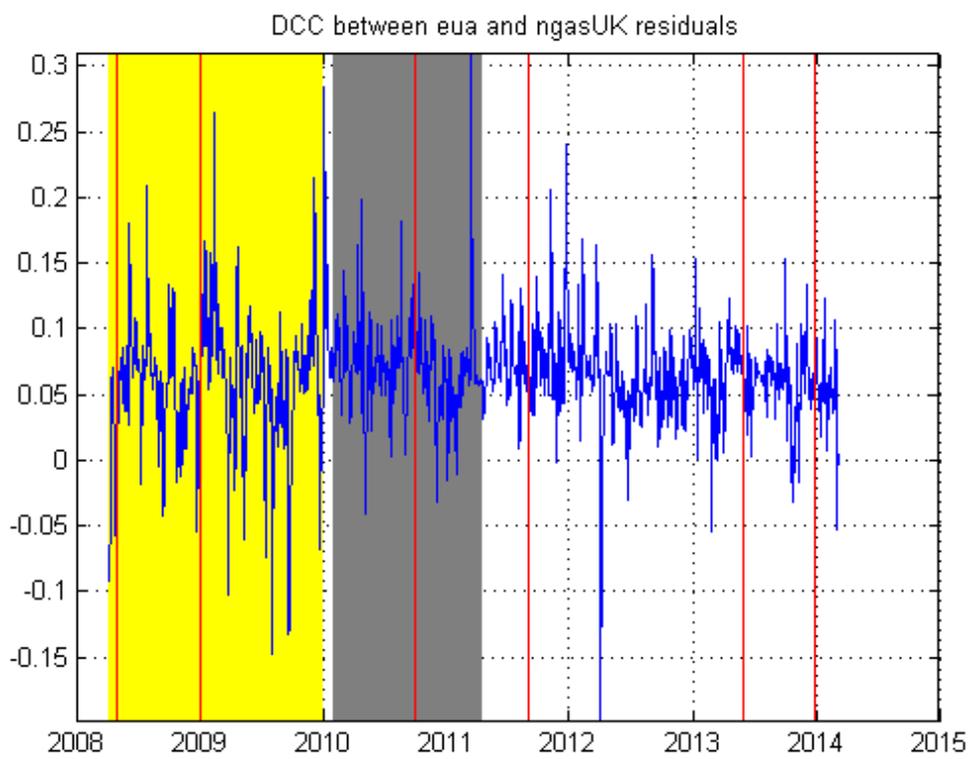

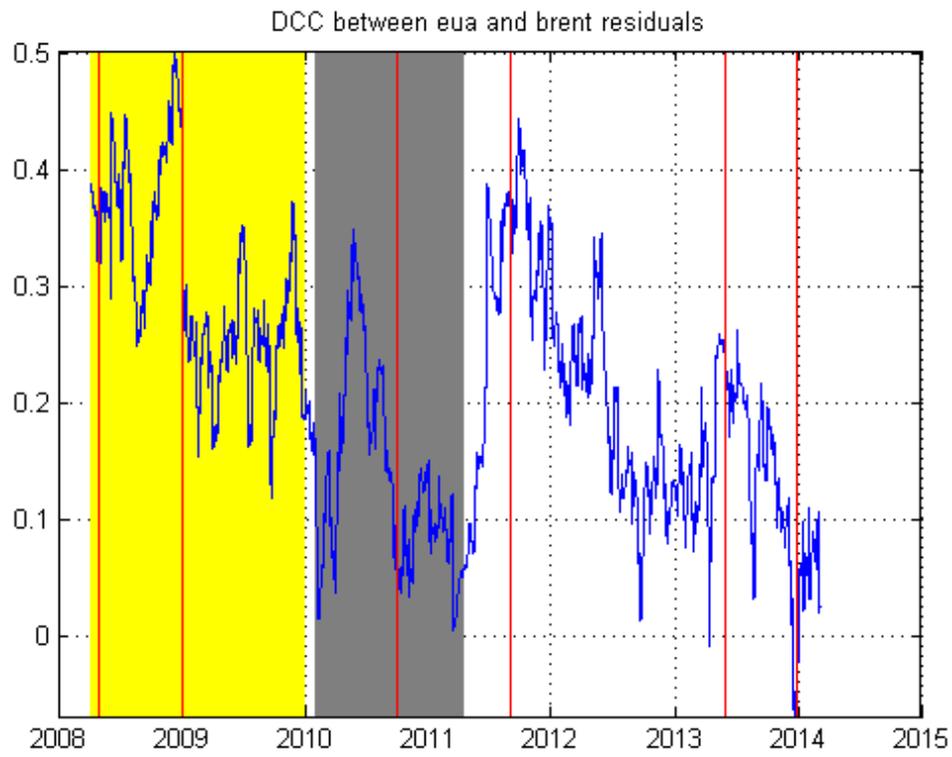

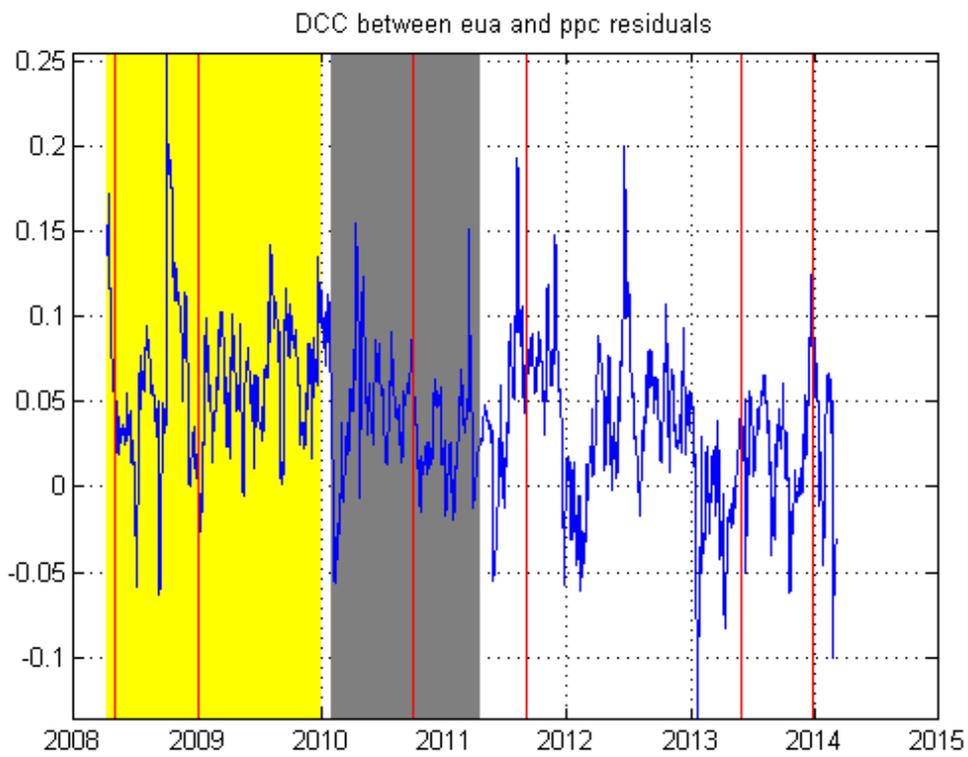

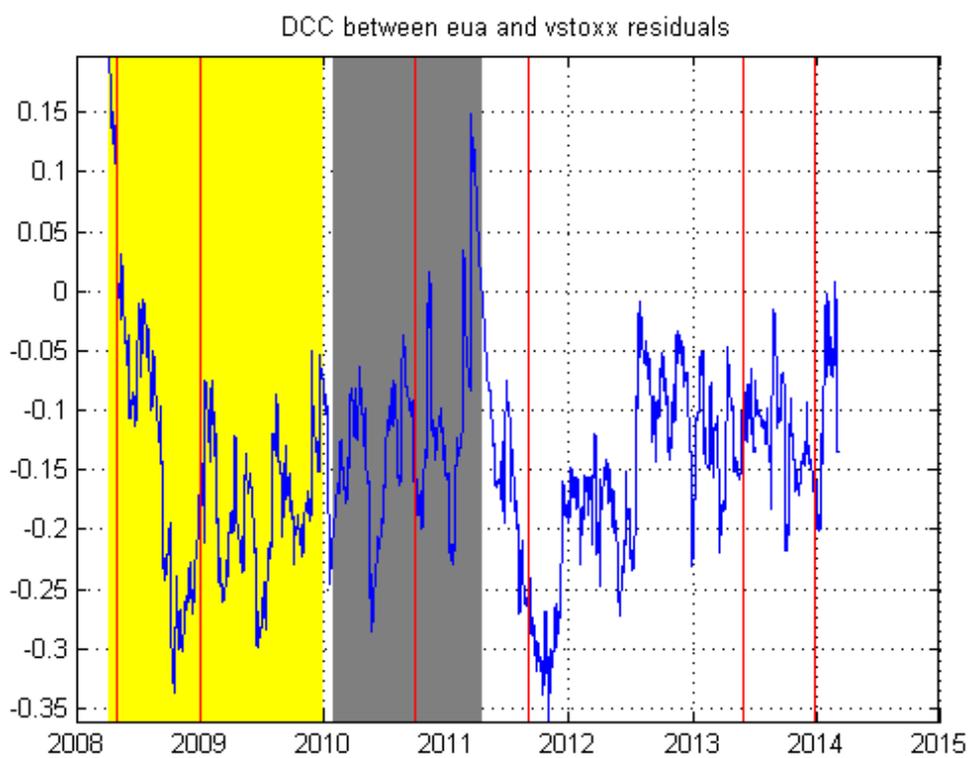

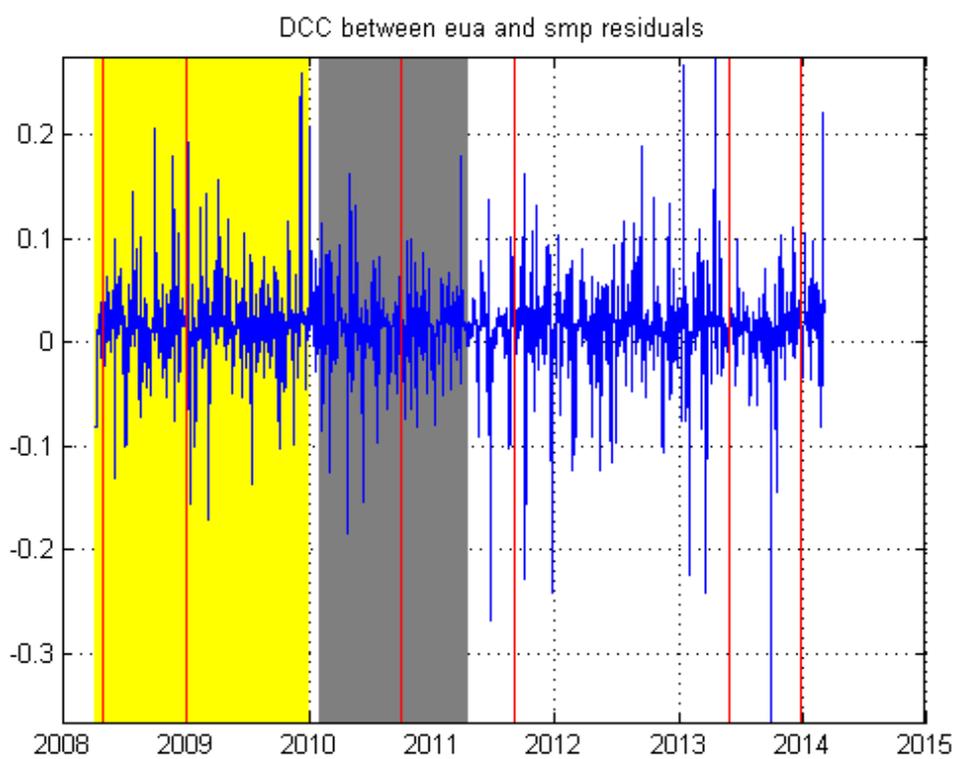

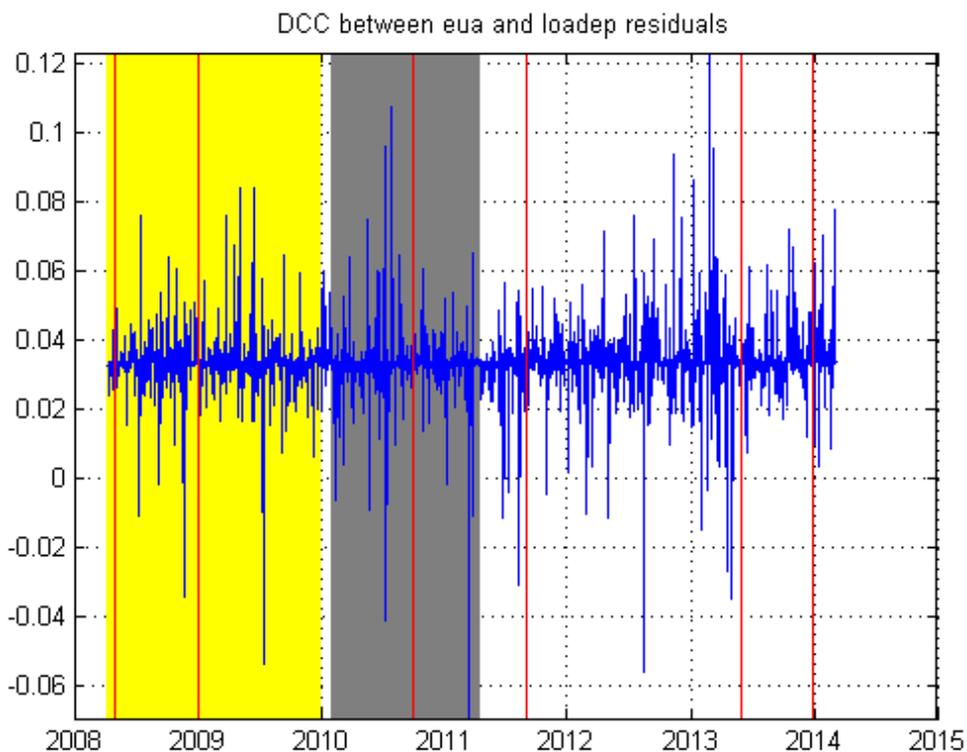

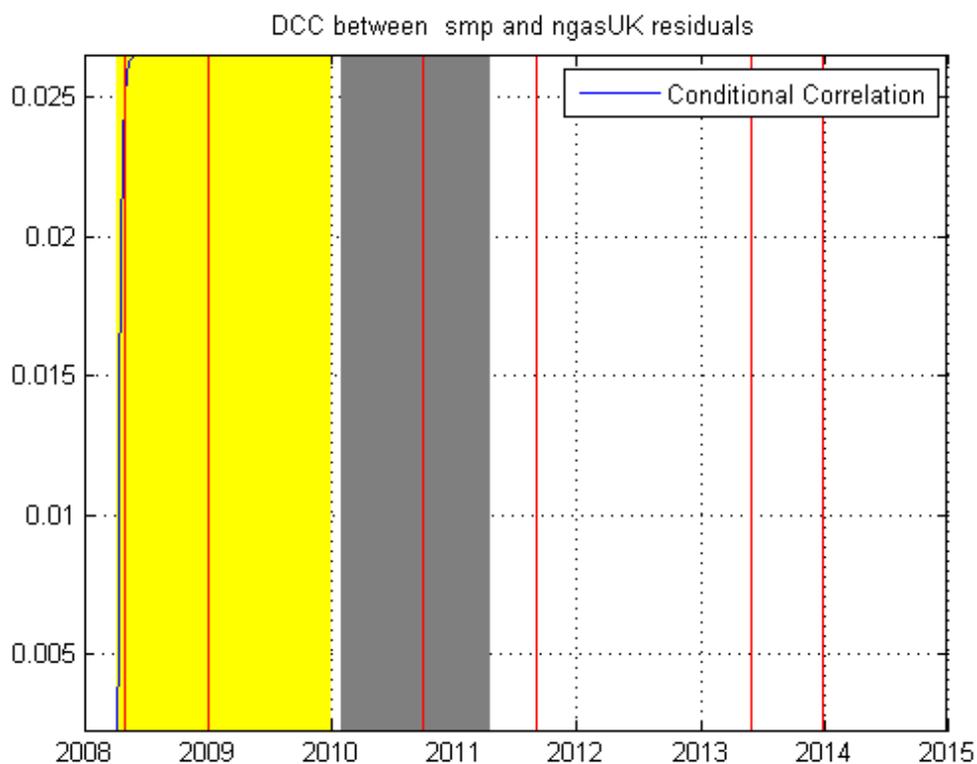

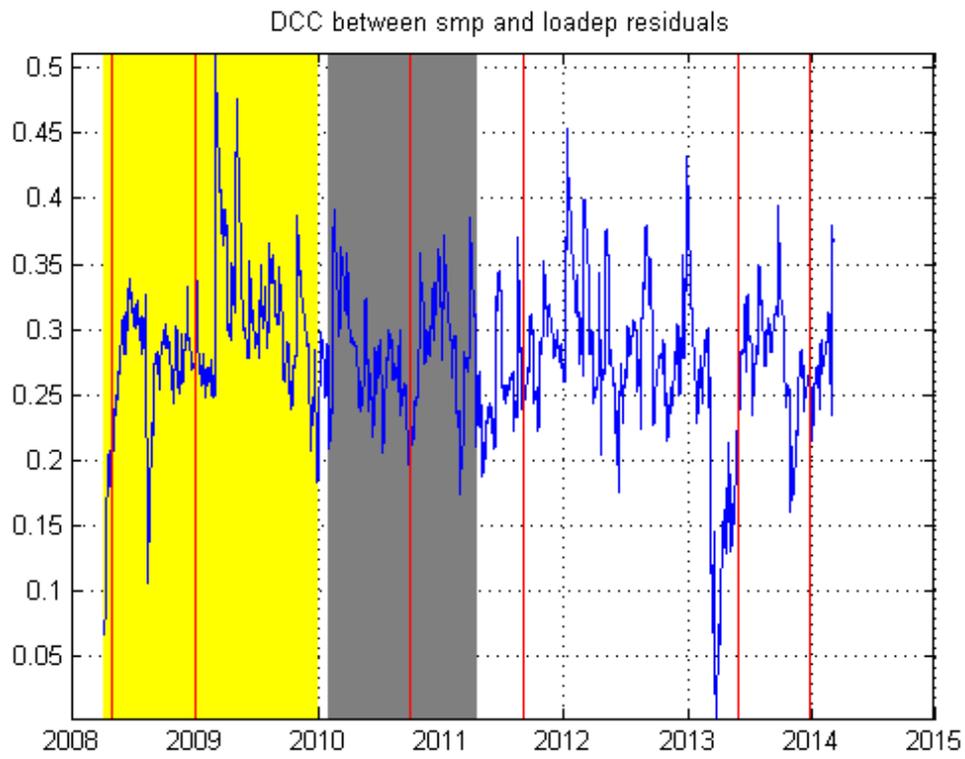

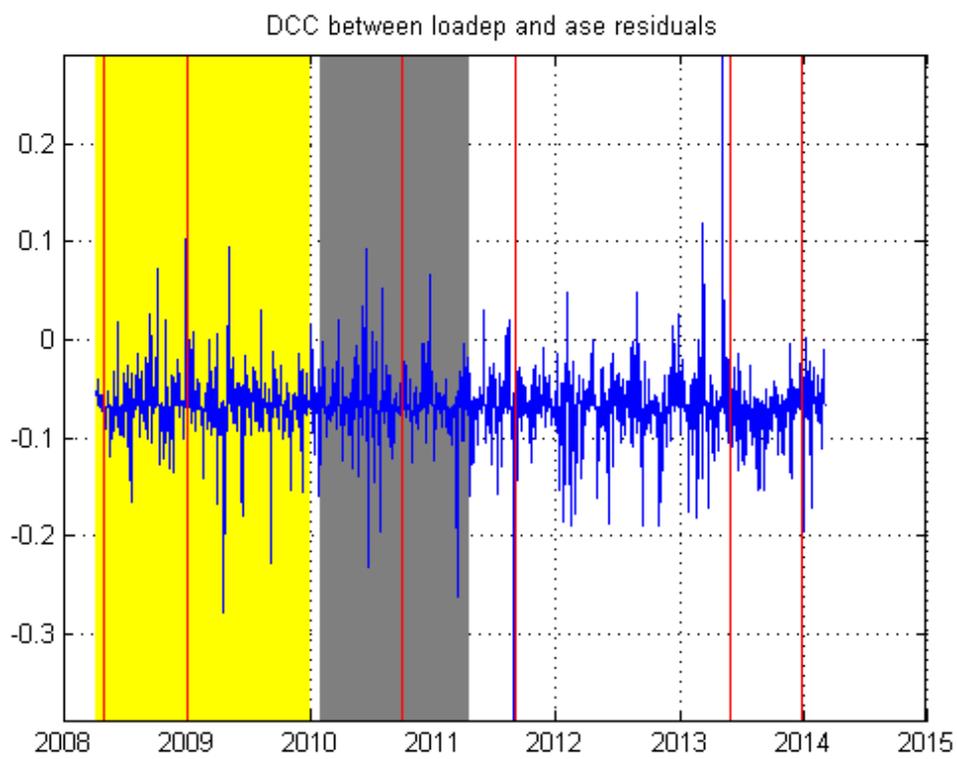

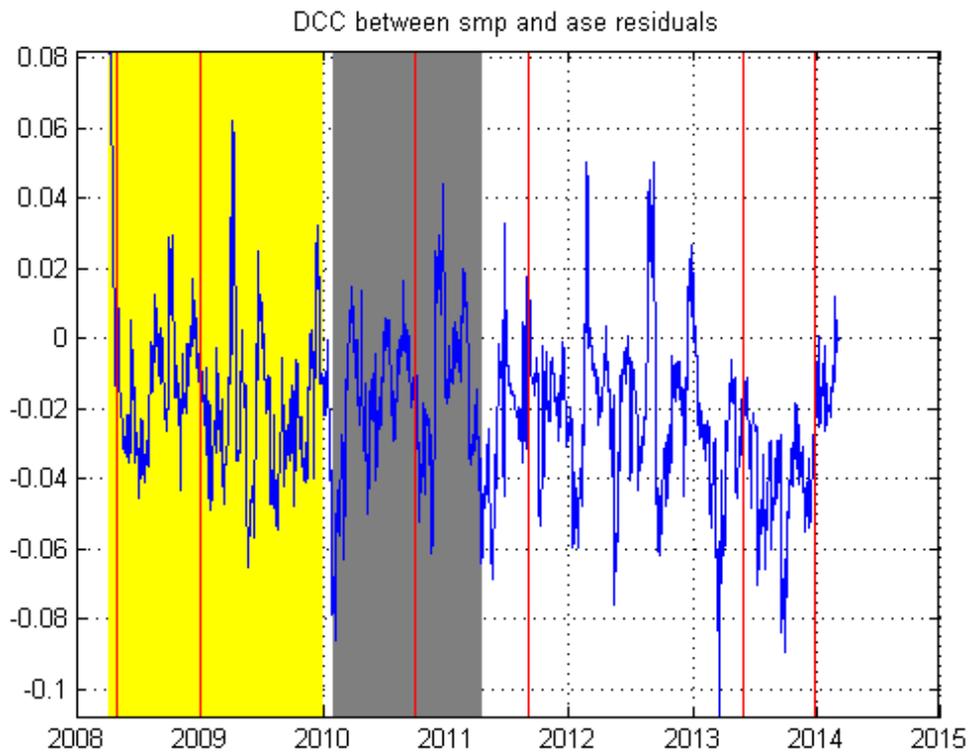

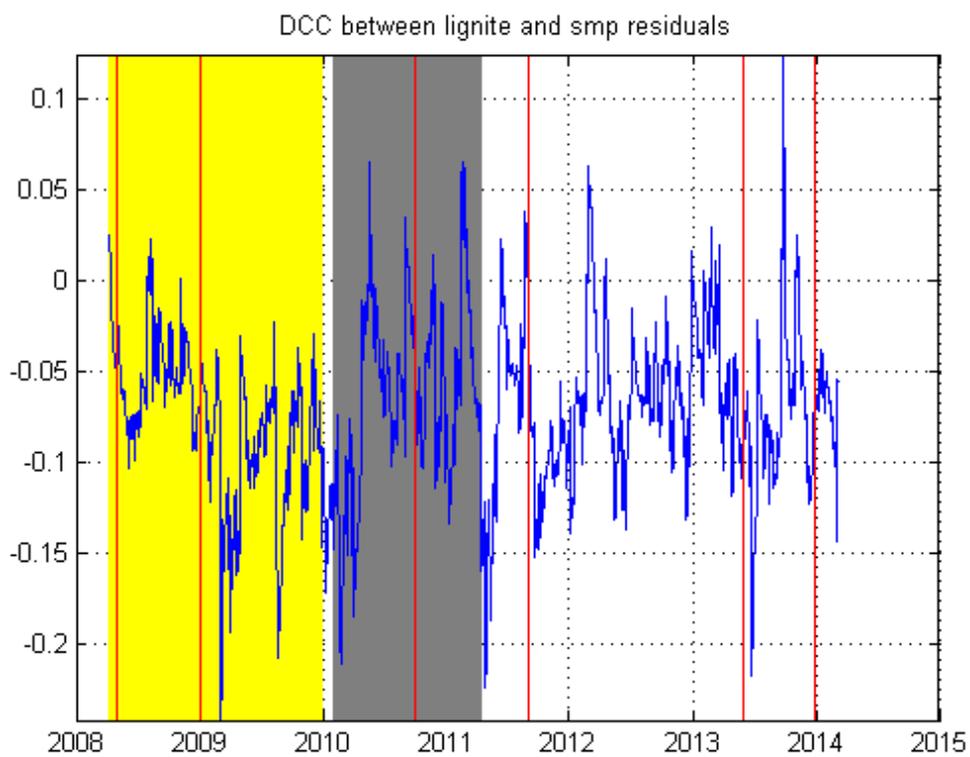

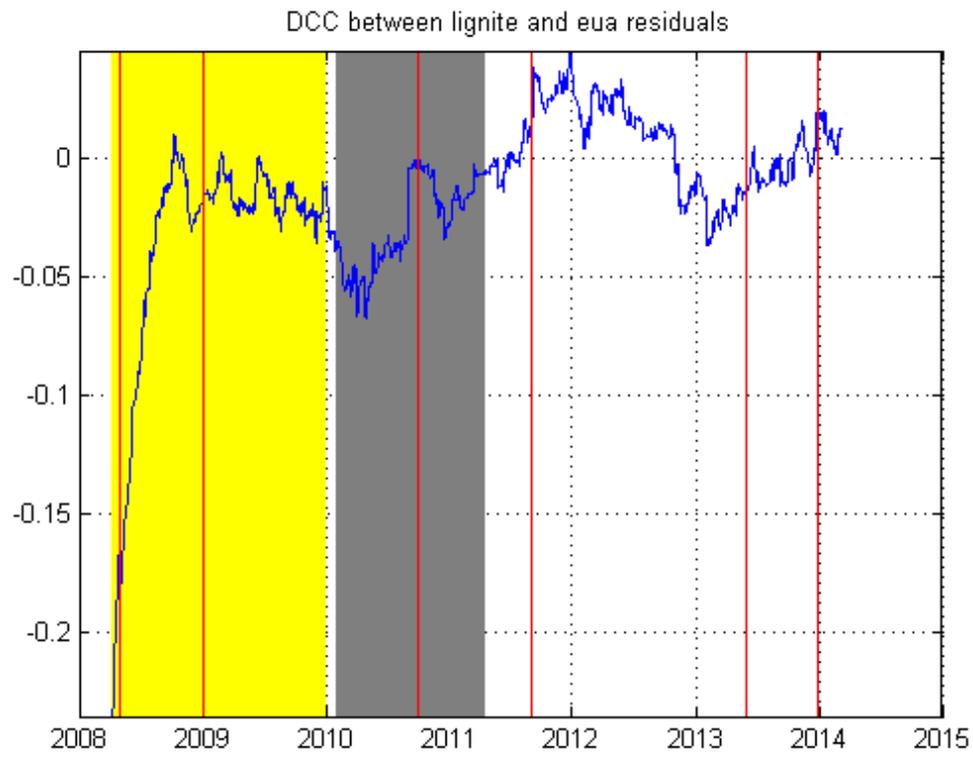

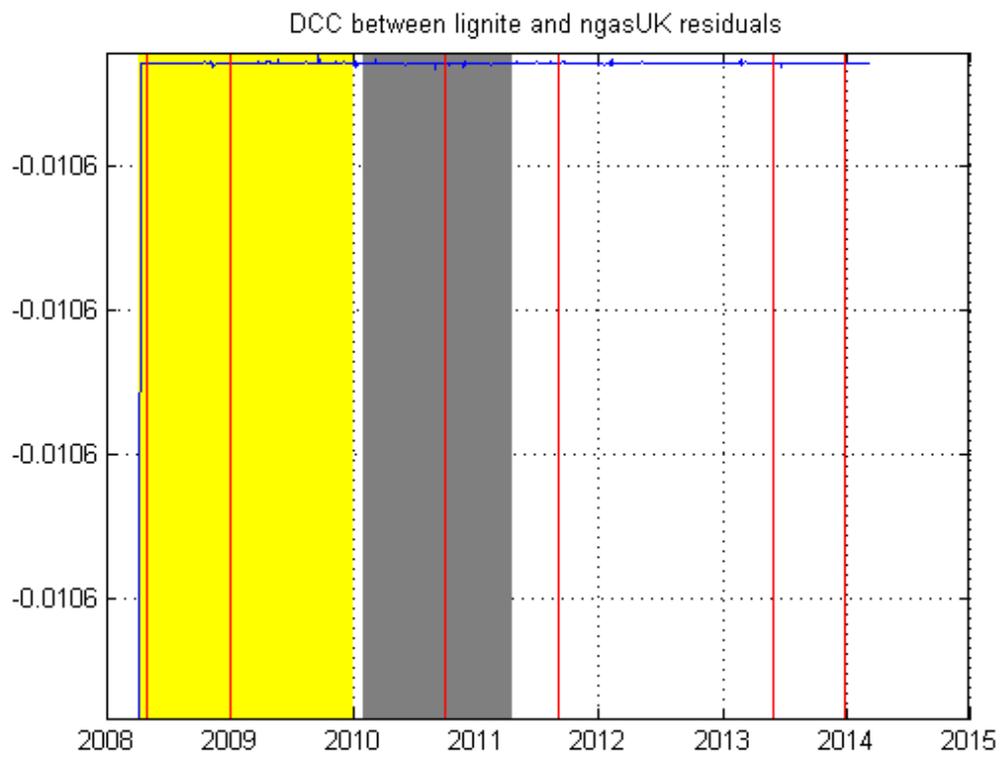

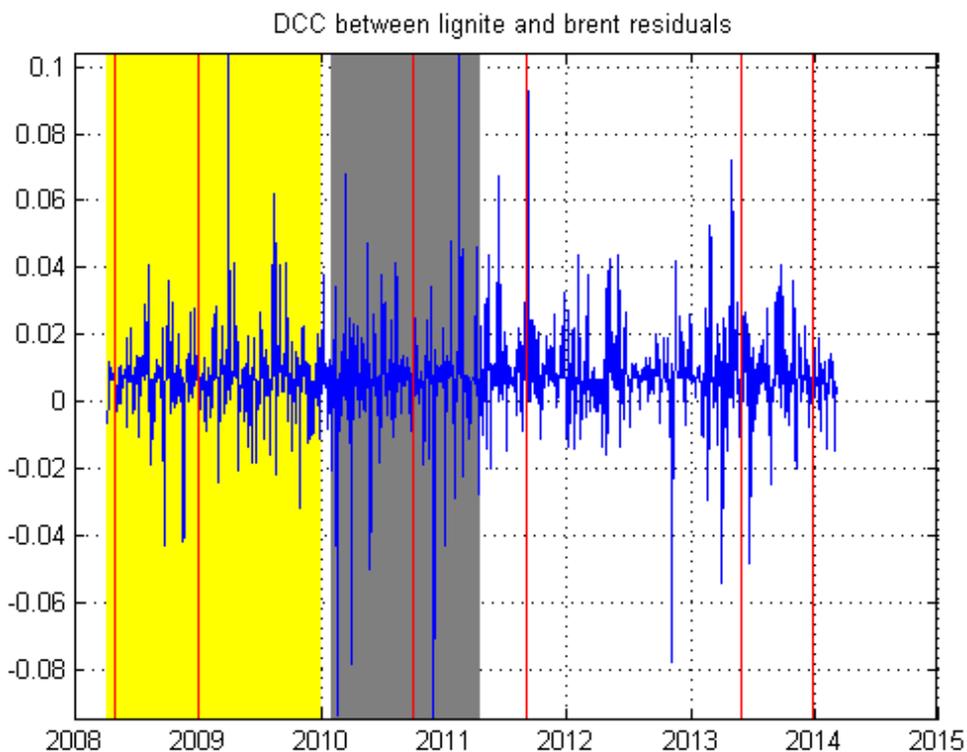

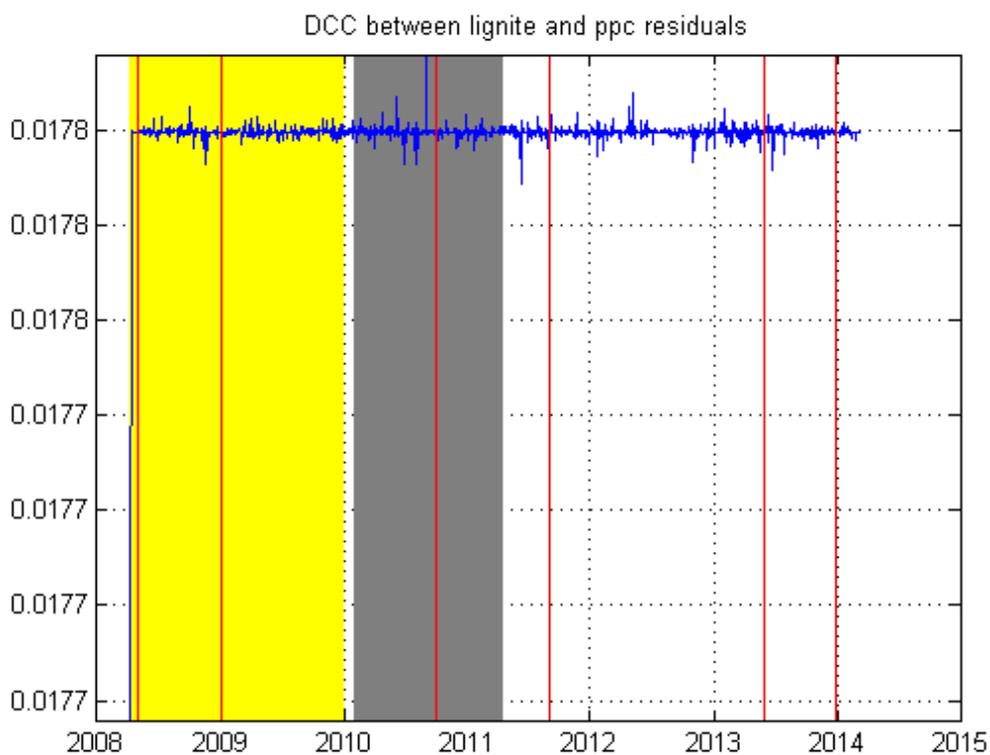

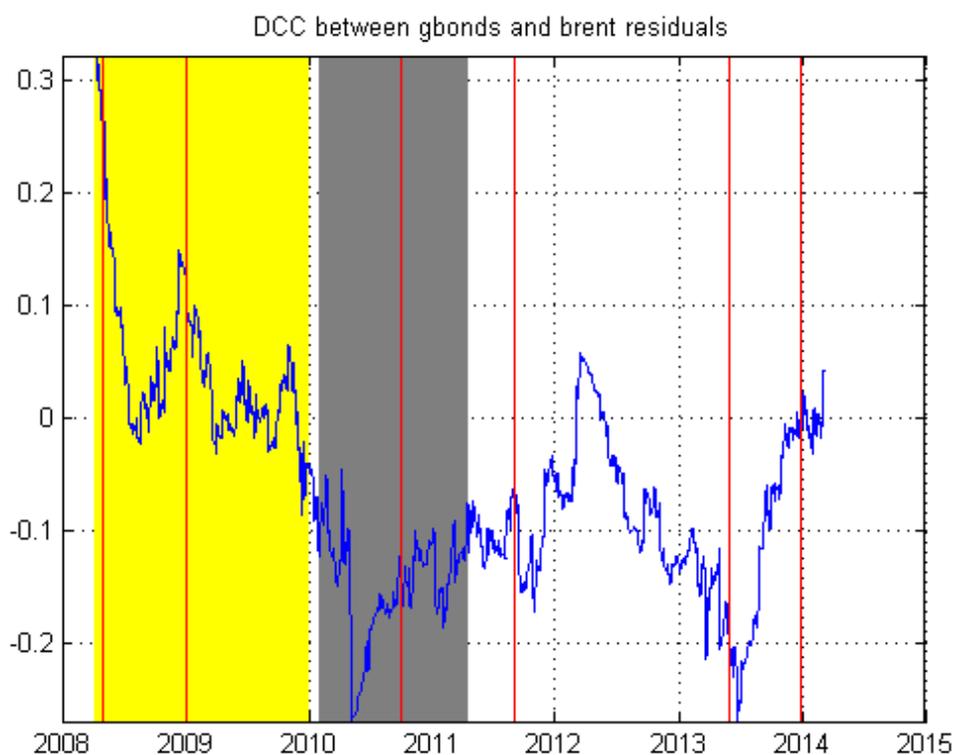

**Table :**

| Year | Expenses (in thousands €) | | Volume (tons) | Annual Average Price (€/ton) |
|---|---|---|---|---|
| | Self-production | 3<sup>rd</sup> parties | | |
| 2006 | 780,000 | 15,000 | | |
| 2007 | 802,000 | 44,000 | | |
| 2008 | 902,000 | 52,000 | | |
| 2009 | 907,000 | 54,000 | | |
| 2010 | 684,300 | 62,300 | | |
| 2011 | 694,900 | 53,600 | | |
| 2012 | 853,500 | 48,400 | | |
| 2013 | 740,500 | 50,200 | | |
| 2014 | 735,800 | 74,400 | | |
| 2015 | 638,000 | 57,600 | | |

# 6. Conclusion

# Acknowledgements

# APPENDIX

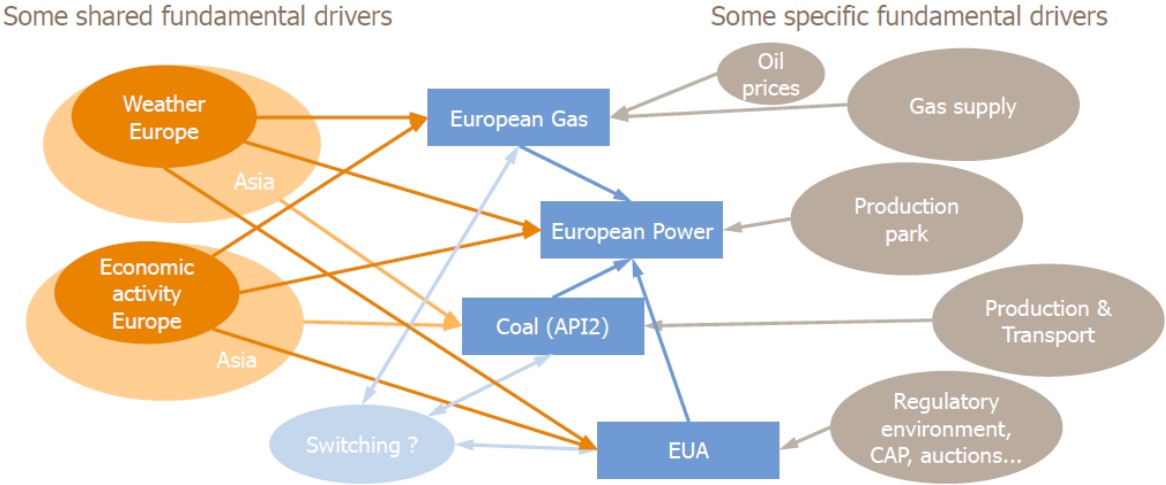